\documentclass[%
 reprint,
superscriptaddress,
preprintnumbers,
nofootinbib,
 amsmath,amssymb,
 aps,
]{revtex4-2}

\usepackage{graphicx}
\usepackage{dcolumn}
\usepackage{bm}
\usepackage{hyperref}
\usepackage{color}

\hypersetup{
    unicode=false,          
    pdftoolbar=true,        
    pdfmenubar=true,        
    pdffitwindow=false,     
    pdfstartview={FitH},    
    pdftitle={},    
    pdfauthor={},     
    pdfsubject={},   
    pdfcreator={},   
    pdfproducer={}, 
    pdfkeywords={} {} {}, 
    pdfnewwindow=true,      
    colorlinks=true,       
    linkcolor=blue, 
    citecolor=blue,        
    filecolor=magenta,      
    urlcolor=blue,           
	breaklinks=true
} 

\usepackage{amssymb}
\usepackage{amsmath}
\usepackage{amsfonts}
\usepackage{xspace}
\usepackage{bm,bbm}
\usepackage{relsize}
\usepackage{siunitx}
\usepackage{xcolor}
\usepackage{upgreek}

\makeatletter
\DeclareFontFamily{OMX}{MnSymbolE}{}
\DeclareSymbolFont{MnLargeSymbols}{OMX}{MnSymbolE}{m}{n}
\SetSymbolFont{MnLargeSymbols}{bold}{OMX}{MnSymbolE}{b}{n}
\DeclareFontShape{OMX}{MnSymbolE}{m}{n}{
    <-6>  MnSymbolE5
   <6-7>  MnSymbolE6
   <7-8>  MnSymbolE7
   <8-9>  MnSymbolE8
   <9-10> MnSymbolE9
  <10-12> MnSymbolE10
  <12->   MnSymbolE12
}{}
\DeclareFontShape{OMX}{MnSymbolE}{b}{n}{
    <-6>  MnSymbolE-Bold5
   <6-7>  MnSymbolE-Bold6
   <7-8>  MnSymbolE-Bold7
   <8-9>  MnSymbolE-Bold8
   <9-10> MnSymbolE-Bold9
  <10-12> MnSymbolE-Bold10
  <12->   MnSymbolE-Bold12
}{}

\let\llangle\@undefined
\let\rrangle\@undefined
\DeclareMathDelimiter{\llangle}{\mathopen}%
                     {MnLargeSymbols}{'164}{MnLargeSymbols}{'164}
\DeclareMathDelimiter{\rrangle}{\mathclose}%
                     {MnLargeSymbols}{'171}{MnLargeSymbols}{'171}
\makeatother

\renewcommand{\vec}[1]{\bm{#1}}

\let\originalleft\left
\let\originalright\right
\renewcommand{\left}{\mathopen{}\mathclose\bgroup\originalleft}
\renewcommand{\right}{\aftergroup\egroup\originalright}

\usepackage{xr}
\makeatletter

\newcommand*{\addFileDependency}[1]{
\typeout{(#1)}
\@addtofilelist{#1}
\IfFileExists{#1}{}{\typeout{No file #1.}}
}\makeatother

\newcommand*{\myexternaldocument}[1]{%
\externaldocument{#1}%
\addFileDependency{#1.tex}%
\addFileDependency{#1.aux}%
}

\myexternaldocument{supp}

\begin{document}

\title{Dynamically reconfigurable topological routing in nonlinear photonic systems}

\author{Stephan Wong}
\email[Email: ]{stewong@sandia.gov}
\affiliation{Center for Integrated Nanotechnologies, Sandia National Laboratories, Albuquerque, New Mexico 87185, USA}

\author{Simon Betzold}
\affiliation{Julius-Maximilians-Universit{\"a}t W{\"u}rzburg, Physikalisches Institut, and W{\"u}rzburg-Dresden Cluster of Excellence ct.qmat, Lehrstuhl f{\"u}r Technische Physik, Am Hubland, W{\"u}rzburg 97074, Germany}
 
\author{Sven H{\"o}fling}
\affiliation{Julius-Maximilians-Universit{\"a}t W{\"u}rzburg, Physikalisches Institut, and W{\"u}rzburg-Dresden Cluster of Excellence ct.qmat, Lehrstuhl f{\"u}r Technische Physik, Am Hubland, W{\"u}rzburg 97074, Germany}

\author{Alexander Cerjan}
\affiliation{Center for Integrated Nanotechnologies, Sandia National Laboratories, Albuquerque, New Mexico 87185, USA}

\date{\today}

\begin{abstract}
The propagation path of topologically protected states is bound to the interface between regions with different topology, and as such, the functionality of linear photonic devices leveraging these states is fixed during fabrication.
Here, we propose a mechanism for dynamic control over a driven dissipative system's local topology, yielding reconfigurable topological interfaces and thus tunable paths for protected routing.
We illustrate our approach in non-resonantly pumped polariton lattices, where the nonlinear interaction between the polaritons and the exciton reservoir due to non-resonant pumping can yield a dynamical change of the topology.
Moreover, using a continuous model of the polariton system based on a driven-dissipative Gross-Pitaevskii equation alongside the spectral localizer framework, we show that the local changes in the nonlinear non-Hermitian system's topology are captured by a local Chern marker.
Looking forward, we anticipate such reconfigurable topological routing will enable the realization of novel classes of topological photonic devices.
\end{abstract}

\maketitle


\section{Introduction}

\begin{figure}[t]
\center
\includegraphics[width=\columnwidth]{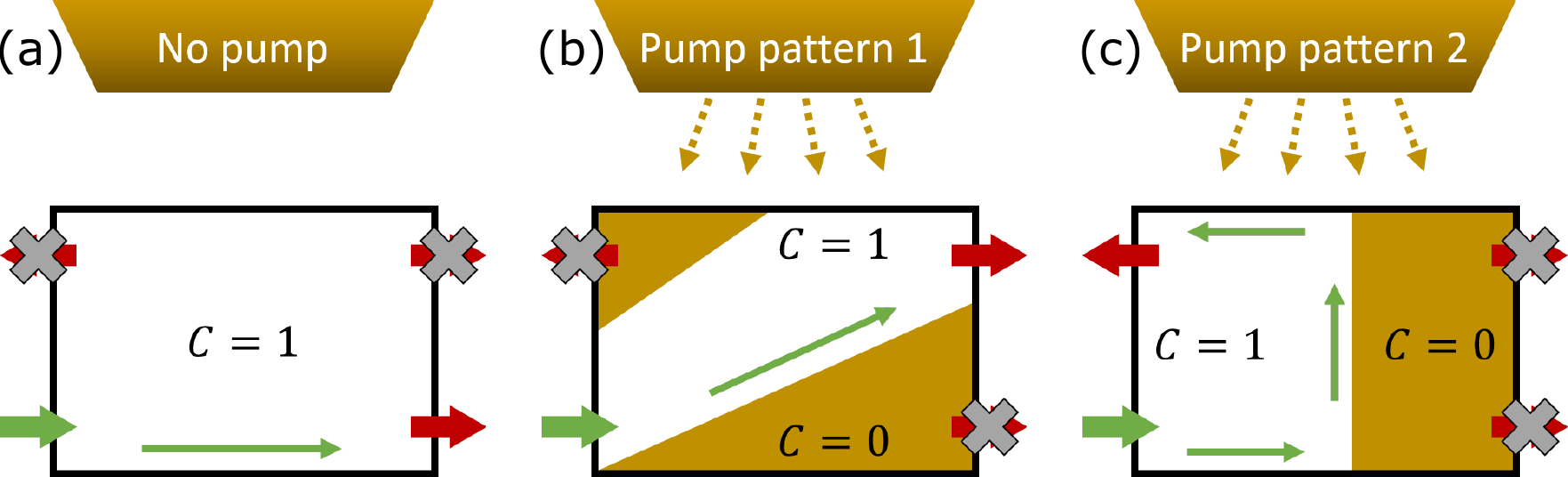}
\caption{
\textbf{Scheme for dynamical control over the topological mode's propagation path.}
(a) Schematic of a topological Chern insulator.
Energy is injected into the chiral edge mode at the green arrow and propagates along the boundary of the lattice.
(b),(c) Illuminating the same nonlinear lattice with different pump patterns (yellow shaded areas) renders the lattice locally topologically trivial and leads to different paths of propagation for the topological mode.
The exact path depends on the shape of the pump pattern, such that the topological mode 
(b) propagates to the top-right port, while avoiding the top-left and bottom-right ports, or 
(c) propagates to the top-left port, while avoiding the top-right and bottom-right ports.
}
\label{fig:scheme}
\end{figure}

Over the past decade, topological photonics has emerged as a promising collection of physical principles for controlling the flow of light.
For example, there is significant interest in harnessing topological phenomena for potential applications and integration to nanophotonic systems for realizing robust photon routing~\cite{Yu2008, He2010, Qiu2011, Fu2011, Peano2016, Lai2016, Ozawa2016, Khanikaev2017, Wu2017, Guglielmon2019, Khanikaev2024} and creating lasers with increased coherence~\cite{Bahari2017, Bandres2018, Dikopoltsev2021, Zeng2020, Amelio2020}.
In particular, the chiral edge modes supported by photonic Chern insulators are an especially enticing class of states for designing next-generation optical devices, as these states both enable non-reciprocal transport and are robust against fabrication imperfections.
Prior studies have observed chiral edge modes in a variety of photonic platforms, including photonic crystals with gyro-optical materials~\cite{Wang2009}, shifted ring-resonator arrays~\cite{Hafezi2011, Hafezi2013}, helical waveguide arrays~\cite{Rechtsman2013}, and more recently exciton-polariton lattices~\cite{Klembt2018}. 

However, the interface localization of a photonic Chern heterostructure's chiral edge modes also presents a substantial limitation on device design: these states' propagation path is fixed by the system's geometry and cannot be readily altered after the device is fabricated.
Although there are previous proposals for reconfigurable topological systems, they are predominantly in the microwave regime with slow reconfigurability time scales in the range of milliseconds~\cite{Cheng2016, Qin2024}, and are thus not technologically relevant integrated nanotechnologies.
At telecommunication wavelengths, prior work on reconfigurable photonics has either been realized through dynamically controlling each lattice element~\cite{Dai2024}, an approach that is challenging to scale, or through tunable spatially non-uniform non-Hermiticity~\cite{Zhao2019}, which requires introducing large material absorptivities.
As such, photonic systems rooted in linear topology are best suited to devices tailored for a single, static function, but are poor candidates for applications requiring dynamic behavior, such as routing.
Nevertheless, while many studies in nonlinear systems have considered topological solitons that can be injected at different lattice sites~\cite{Leykam2016, Zhou2017, Maczewsky2020, Smirnova2020, Mukherjee2021, Jorg2024}, these states are still constrained along the lattice's structural boundary or remain confined in the bulk~\cite{Pieczarka2021, Mandal2023, Lackner2024}.
Altogether, an ideal reconfigurable topological platform would exhibit a scalable, fast dynamic method for changing the system's local Chern phase without introducing additional propagation losses, so that a reconfigurable router can take full advantage of a chiral edge mode's inherent reflectionless propagation~\cite{Ozawa2019} and robustness against dephasing~\cite{Karcher2024}.

Here, we propose an approach for realizing fast reconfigurable local Chern topology in nonlinear photonic systems and numerically illustrate our method in a non-resonantly pumped exciton-polariton platform~\cite{Solnyshkov2021} operating at telecommunication wavelengths. 
In particular, the part-photon, part-exciton nature of the system's polaritons yields a local nonlinear matter-matter interaction with the available excitons, which form a dissipative reservoir that can be populated through a non-resonant pump.
Thus, starting with a topologically non-trivial polariton lattice~\cite{Karzig2015, Bardyn2015, Nalitov2015, Yi2016, Klembt2018, Solnyshkov2021} [Fig.~\ref{fig:scheme}(a)], the incident pump can locally change the system's topology to be trivial, creating a topological interface within the system whose boundary is based on the pump pattern. 
As such, any incident signal propagating in a chiral edge state of the topological portion of the lattice is guaranteed to follow the new topological boundary [Figs.~\ref{fig:scheme}(b),(c)]. 
Moreover, as the exciton reservoir is dissipative, if the non-resonant pump is turned off the full lattice returns to its original topological phase after a characteristic decay time, but this dissipation does not strongly influence the polaritons such that their propagation remains relatively loss-less.
Additionally, superposing the reconfigurable Chern topology features using multiple non-resonant pump patterns and pump amplitudes enables reconfigurable multi-channel topologically protected routing via different topological interfaces in different energy ranges.
We use a continuum model of the polariton system based on the driven-dissipative Gross-Pitaevskii equations for experimentally realizable parameters~\cite{Klembt2018} and spatially resolve the change in the pumped system's local Chern number using a nonlinear non-Hermitian generalization of spectral localizer framework~\cite{Loring2015, Loring2017, Loring2020, Cerjan2023, Dixon2023, Wong2024}.
Given the possible tiny extent of the pumped region, the spectral localizer provides a rigorous and quantitative understanding of the local topology, as opposed to topological band theory predicated on having a sufficiently large uniform system.
Overall, while there has been growing interest in inducing topological phase changes via optical pumping without relying on external magnetic fields~\cite{Bleu2016, Bleu2017, Sigurdsson2017, Solnyshkov2018, Sigurdsson2019, Zheng2023, Solnyshkov2021, Banerjee2021, Banerjee2020, Ma2020}, our results show that it is possible to dynamically reconfigure a system's topological interfaces post-fabrication by leveraging a nonlinear response, a phenomena that should be available in a variety of systems featuring nonlinear interactions, and may be of practical use for multitasking devices while retaining the robustness of topological protected edge modes.
%


\section{Results}


\subsection{Nonlinearly-induced topological interface}

To illustrate dynamic control of local topology, we consider Chern polariton lattices consisting of quantum wells embedded in a honeycomb array of vertical microcavities~\cite{Weisbuch1992, Klembt2018} under non-resonant pumping.
In particular, the non-resonant pumping populates an exciton reservoir, and the dynamics of the polaritons $\vec{\psi}(\vec{x},t)$ with the exciton reservoir $n_{\textrm{r}}(\vec{x},t)$ are given by a driven-dissipative Gross-Pitaevskii equation~\cite{Wouters2007, Banerjee2021}
\begin{align}
\label{eq:polariton_rate}
\begin{split}
i \hbar \frac{\partial}{\partial t} \vec{\psi} 
& = H_0 \vec{\psi} - i \hbar \frac{\gamma_c}{2} \vec{\psi} + g_c \left| \vec{\psi} \right| ^2 \vec{\psi} \\
& \qquad \qquad + \left( g_r + i \hbar \frac{R}{2} \right) n_r \vec{\psi} + S_{\text{probe},},
\end{split}
\\
\label{eq:reservoir_rate}
\frac{\partial}{\partial t} n_{r} 
& = - \left( \gamma_r + R \left| \vec{\psi}  \right| ^2 \right) n_{r}  + S_{\text{pump}}
.
\end{align}
Here,
$H_0$ is the system's Hamiltonian, 
$\gamma_c$ and $\gamma_r$ are the relaxation rates for the polariton state and exciton reservoir, 
$g_c$ and $g_r$ are the polariton-polariton and polariton-exciton interaction strengths, 
$R$ is the amplification rate of the polariton state due to stimulated scattering from the reservoir,
$S_\text{probe}(\vec{x}, t)$ is the resonant probe for directly exciting the polaritons, and $S_\text{pump}(\vec{x}, t)$ is the non-resonant pump used for injecting free carriers.
When $S_\text{pump}(\vec{x}, t) = 0$, the non-trivial topology in the polariton lattice~\cite{Karzig2015, Bardyn2015, Nalitov2015, Yi2016, Klembt2018, Solnyshkov2021} results from the interplay of an external magnetic field and an effective spin-orbit coupling.
The external magnetic field breaks time-reversal symmetry, and induces a Zeeman splitting between the spin-up and spin-down excitons, while the effective spin-orbit coupling originating from the coupling between the transverse electric (TE) and magnetic (TM) photonic modes opens a topological gap by coupling the two spin sectors.
The corresponding Hamiltonian $H_0$ is given, in the polariton spin basis $\vec{\psi}(\vec{x}) = [\vec{\psi}_+(\vec{x}), \vec{\psi}_-(\vec{x})]$, by
\begin{equation}
\label{eq:H0}
\begin{split}
& H_0 = \\
& 
\left(
\begin{array}{cc}
-\frac{\hbar^2}{2m} \nabla^2 + V(\vec{x}) + \frac{1}{2} \Delta_\text{eff} & -\beta_\text{eff} \left( \partial_x - i\partial_y \right)^2 \\[1.5ex]
-\beta_\text{eff} \left( \partial_x + i\partial_y \right)^2 & -\frac{\hbar^2}{2m} \nabla^2 + V(\vec{x}) - \frac{1}{2} \Delta_\text{eff}\\
\end{array}
\right)
,
\end{split}
\end{equation}
where 
$m$ is the polariton mass,
$V(\vec{x})$ is the polariton potential patterned as a honeycomb lattice,
$\beta_\text{eff}$ is the spin-orbit coupling strength, and
$\Delta_\text{eff}$ is the Zeeman coefficient.
Altogether, the instantaneous polariton Hamiltonian $H$ is thus non-Hermitian and nonlinear,
\begin{equation}
\label{eq:H}
H(t,\vec{\psi}, n_{\textrm{r}}) = H_0 + i \Gamma(\psi, n_r) + B(\psi, n_r)
,
\end{equation}
where
$\Gamma(\psi, n_r) = - \hbar \frac{\gamma_c}{2} + \hbar \frac{R}{2} n_r$ gathers the non-Hermitian terms in Eq.~\eqref{eq:polariton_rate} from the driven-dissipative system, and
$B(\psi, n_r) = g_c \left| \vec{\psi} \right| ^2 + g_r n_r$ is the blueshift arising from the repulsive nonlinear interaction between the polaritons and the excitonic reservoir~\cite{Kasprzak2006, Wouters2007, Wertz2010, Liu2015, Panico2023}. 

A polariton system's topological dynamics can be identified directly from the continuous Hamiltonian model [Eq.~\eqref{eq:H}] of a finite structure using the spectral localizer framework, which enables local topological classification at a specified location and energy for a nonlinear system with a given occupation~\cite{Wong2023, Bai2024}.
In particular, we generalize the two-dimensional (2D) spectral localizer for static non-Hermitian systems~\cite{Cerjan2023, Dixon2023} to incorporate nonlinearities, such that the instantaneous nonlinear non-Hermitian spectral localizer $L_{(\vec{x},E)}$ is
\begin{align}
\label{eq:localizer}
\begin{split}
& L_{(\vec{x},E)}(X, Y, H(t,\vec{\psi},n_{\textrm{r}})) = \\[1.ex]
& 
\left(
\begin{array}{cc}
\left[H(t,\vec{\psi},n_{\textrm{r}}) - E \mathbf{1} \right] & \kappa (X - x \mathbf{1}) - i\kappa(Y - y \mathbf{1}) \\[1.ex] 
\kappa (X - x \mathbf{1}) + i\kappa(Y - y \mathbf{1}) & -\left[H(t,\vec{\psi},n_{\textrm{r}}) - E \mathbf{1} \right]^\dagger
\end{array}
\right)
,
\end{split}
\end{align}
where
$H$ is the Hamiltonian matrix derived from the finite-difference discretization of the instantaneous continuous model [Eq.~\eqref{eq:H}],
$X$ and $Y$ are the position matrices that (in this basis) are diagonal with entries corresponding to the real-space coordinates of the finite-difference degrees-of-freedom in the $x$- and $y$-directions, and
$\mathbf{1}$ is the identity matrix.
In equation~\ref{eq:localizer}, the subscript $(\vec{x},E)$ indicates the location $(\vec{x})$ and energy $(E)$ where the local topology will be classified, and $\kappa$ is a scaling coefficient that enforces consistent units between the position and Hamiltonian matrices. $\kappa$ also ensures balanced spectral weights on the system's position information relative to its Hamiltonian, and is typically of the order $\kappa \sim E_\text{gap} / L$~\cite{Loring2020, Dixon2023, Cerjan2024} where $E_\text{gap}$ is the relevant spectral gap and $L$ the length of the finite system considered.
Using $L_{(\vec{x},E)}$, instantaneous local topology at some spatial-energy coordinate $(\vec{x},E)$ can be determined using the local Chern number $C_{(\vec{x},E)}^{\textrm{L}}$~\cite{Loring2015, Cerjan2023}
\begin{multline}
\label{eq:local_chern_nb}
C_{(\vec{x},E)}^{\textrm{L}}(X, Y, H(t,\vec{\psi},n_{\textrm{r}})) = \\
\frac{1}{2} \textrm{sig}_\mathbb{R} \left[ L_{(\vec{x},E)}(X, Y, H(t,\vec{\psi},n_{\textrm{r}})) \right]
,
\end{multline}
where $\textrm{sig}_\mathbb{R}(M)$ is the signature of the line-gapped matrix $M$, i.e., the difference between the number of eigenvalues with positive and negative real parts. Note, $C_{(\vec{x},E)}^{\textrm{L}}$ is provably equal to the global Chern number for crystalline gapped systems with $E$ chosen in the relevant band gap~\cite{Loring2020}.
Altogether, the topology of the exciton-polariton lattice can be identified using the local Chern number $C_{(\vec{x}_0,E)}^{\textrm{L}}$ with $\vec{x}_0$ chosen inside the system's bulk and $E$ the given energy of interest [see Sect.~\ref{sect_supp:localizer} for further details~\cite{supp}]. 
Specifically, the spectral localizer offers a rigorous and quantitative understanding of the topology over small local regions, contrary to topological band theory which is predicated on having sufficient large uniform system.

\begin{figure}[!]
\center
\includegraphics[width=\columnwidth]{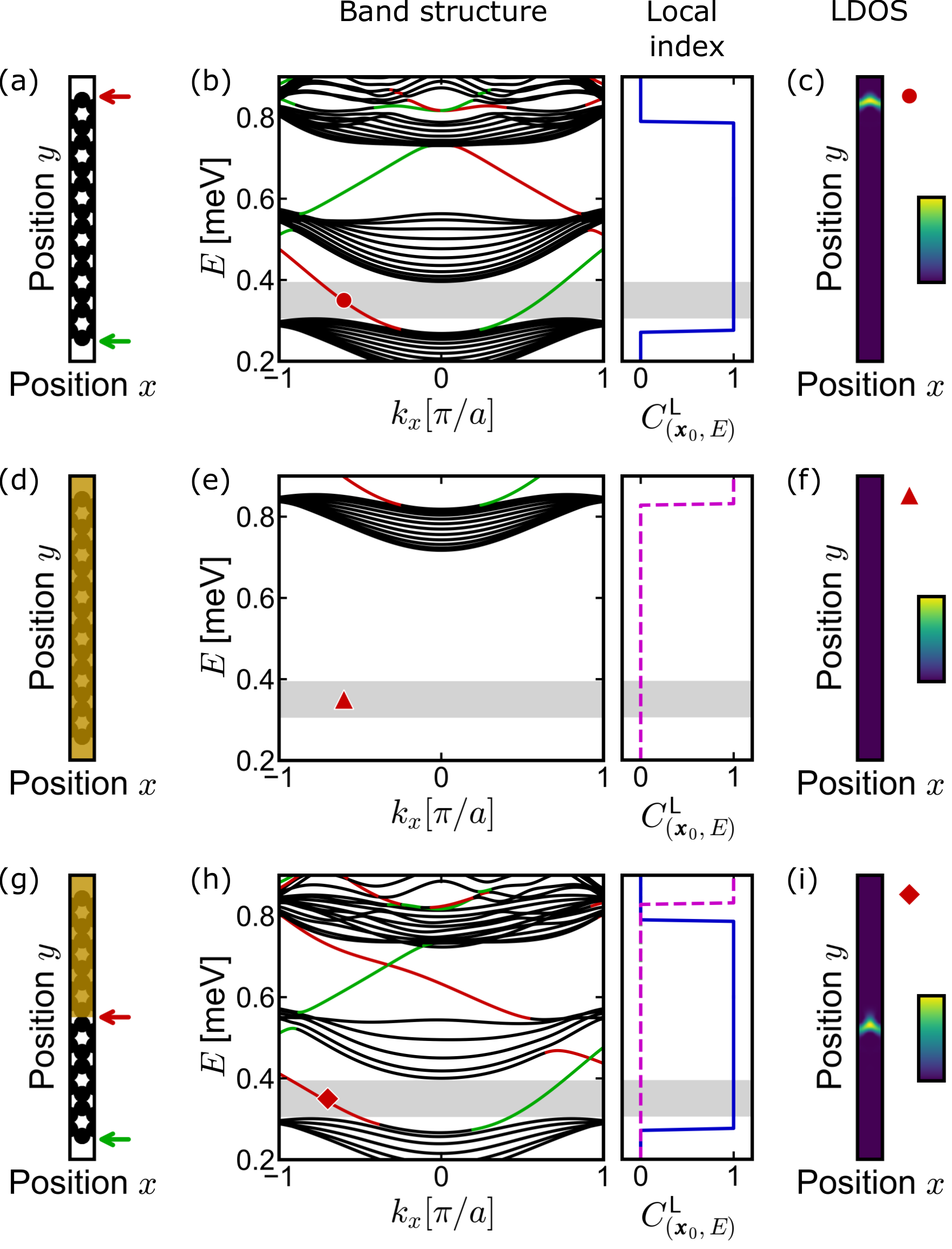}
\caption{
\textbf{Change of the topology due to reservoir-induced blueshift.}
(a) Potential landscape $V(\vec{x})$ of the ribbon polariton lattice arranged in a honeycomb lattice. 
The black and white regions correspond to potentials of $V=6~\unit{\meV}$ and $V=0~\unit{\meV}$, respectively.
(b) Ribbon band structure and the corresponding local Chern number $C_{(\vec{x}_0,E)}^{\textrm{L}}$ calculated using the spectral localizer.
In the band structure, the black lines correspond to bulk modes, and the green (red) lines denote the chiral edge mode dispersion localized at the bottom (top) side of the lattice, as shown by the color-coded arrows in (a).
The gray shaded area indicates the energy range of interest. 
(c) Local density of states (LDOS) of the red line at $E=0.35~\unit{\meV}$, $k_x = -0.6 [\pi/a]$. 
(d)-(f) Same as (a)-(c) but with an yellow overlay depicting the blueshift of the whole ribbon structure.
(g)-(i) Same as (a)-(c) but with the blueshift applied only to half of the ribbon lattice. 
The red (and green) lines in the band structure correspond to the topological edge modes localized at the interface between the blueshifted and non-blueshifted areas (and at the bottom edge of the lattice).
In (h), the local Chern number obtained from both the pumped (dashed magenta line) and unpumped (solid blue line) regions are shown, demonstrating the different topology between the pumpped and unpumped region at the given energy range of interest (gray shaded area).
(i) LDOS of the red line at $E=0.35~\unit{\meV}$, $k_x = -0.7 [\pi/a]$.
Simulations use a reservoir-induced blueshift of $E_\text{blueshift}=0.55~\unit{\meV}$ and spectral localizer calculations use $\kappa = 0.015~\unit{\meV\per\um}$ and a finite 2D system of size $26.1~\unit{\um} \times 23.7~\unit{\um}$. 
}
\label{fig:band}
\end{figure}

To demonstrate how the nonlinearity inherent in exciton-polariton system can yield a shift in the lattice's local topology, we consider a ribbon geometry of the 2D system, i.e., periodic in one direction and finite in the other [Fig.~\ref{fig:band}(a)], and use experimentally realizable parameters from Ref.~\cite{Klembt2018}. 
The ribbon band structure of the unpumped system [Fig.~\ref{fig:band}(b)] features a topologically non-trivial band gap [gray shaded area] with corresponding chiral edge modes [see red/green solid lines].
The right panel of Figure~\ref{fig:band}(b) indicates the local Chern number as a function of energy, confirming the unpumped system's topology. 
As the entire lattice is non-trivial, there is a topological interface at the lattice's boundaries and therefore the chiral edge modes can be identified in the local density of states (LDOS) at the structure's edges [Fig.~\ref{fig:band}(c)].

However, by including a reservoir-induced blueshift, the topology at the given energy range of interest can change.
For example, by non-resonantly pumping on the entire lattice, the polariton potential landscape blueshifts [Fig.~\ref{fig:band}(d)] and thus the system's band structure does as well [Fig.~\ref{fig:band}(e)].
As such, this blueshift induces the lattice's topology to become trivial in the chosen energy range [gray shaded area in Fig.~\ref{fig:band}(e)].
Therefore, if only half of the lattice is non-resonantly pumped [Figs.~\ref{fig:band}(g),(h)], only that portion of the lattice will be rendered trivial in the energy range of interest.
In other words, such non-uniform pumping creates a topological interface between the pumped and unpumped regions within the lattice.
Moreover, the corresponding ribbon band structure for this nonlinearly induced ``heterostructure'' reveals the existence of chiral edge modes that cross the bulk band gap at the chosen energy [Fig.~\ref{fig:band}(h)] that correspond to states that are localized at the newly formed topological interface [Fig.~\ref{fig:band}(i)] and at the lattice's boundary.
Notably, the topological transition across the ``heterostructure'' interface is attributed to a reservoir-induced blueshift, where instead of changing the overall topology of the exciton-polariton manifold, the repulsive nonlinear interaction between the exciton reservoir and the exciton-polariton will shift the energies of the polariton states relative to the unpumped regions.


\subsection{Topological routing with non-resonant pump patterns}
\label{sect:reconfig}

\begin{figure*}[!]
\center
\includegraphics[width=2\columnwidth]{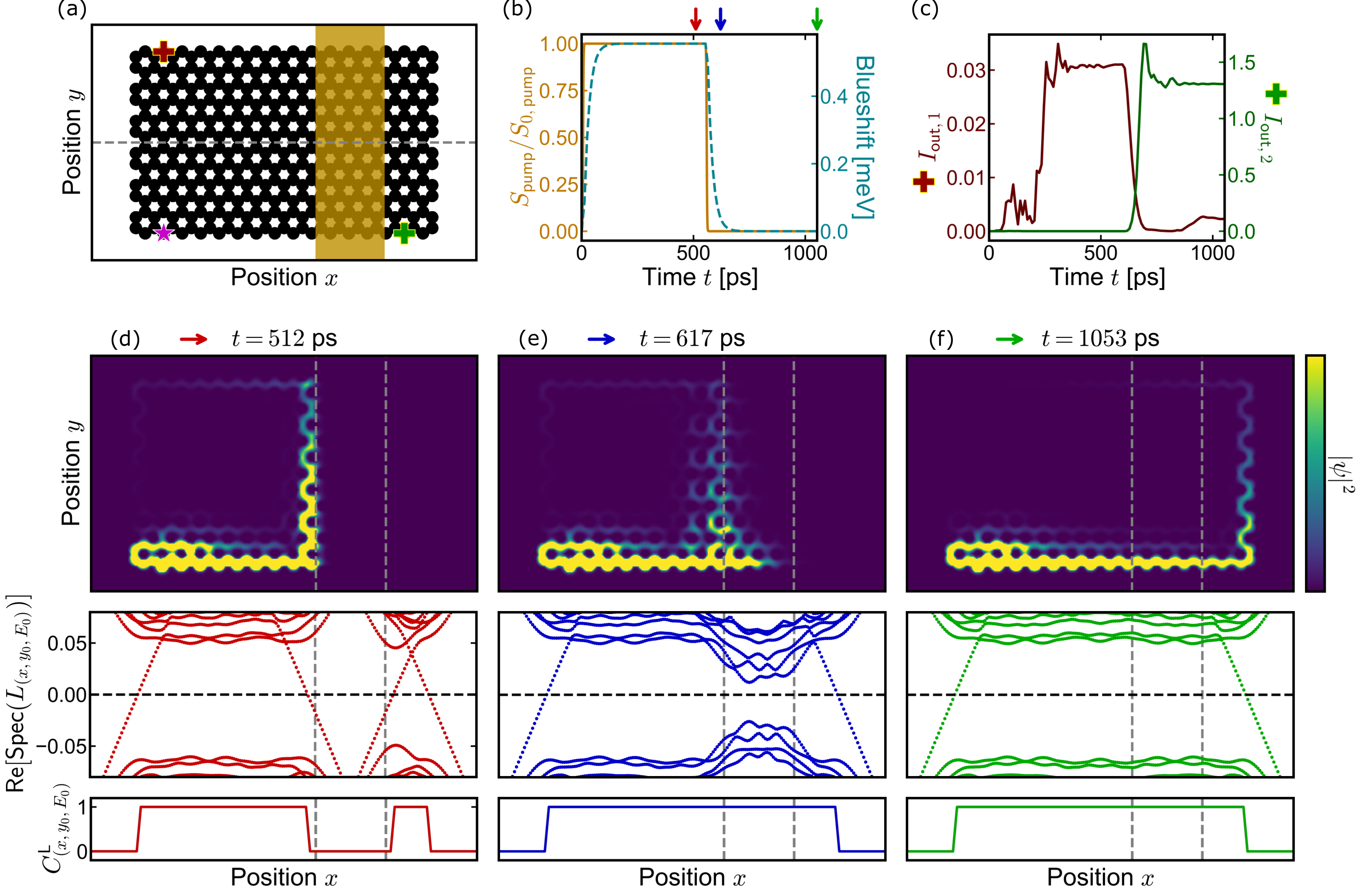}
\caption{
\textbf{Reconfigurable topological routing with non-resonant pumping.}
(a) Potential landscape $V(x,y)$ of the polariton lattice arrange in a honeycomb lattice.
The yellow shaded area depicts the pump pattern, the magenta star indicate the position of the probe source.
(b) Temporal evolution of the non-resonant pump power (cyan solid line) and the induced blueshift (blue dashed line).
(c) Output intensity integrated over a width of $2~\unit{\um}$ in the y-direction, at the positions given by the dark red and dark green crosses in (a).
(d)-(f) Snapshot of the total intensity of the polariton $|\psi|^2 = |\psi_+|^2 + |\psi_-|^2$ at the time indicated by the color-coded arrows in (b), with the corresponding eigenvalues of the spectral localizer $\text{Re} \left[ \text{Spec} \left( L_{(x,y_0,E_0)} \right) \right]$ and the local Chern number $C_{(x,y_0,E_0)}^{\textrm{L}}$ along the gray dashed line in (a) and at $E_0 = 0.35~\unit{\meV}$.
The parameter values for the Hamiltonian are the same as in Fig.~\ref{fig:band}.
Dynamical parameter values:
pump amplitude $S_{0,\text{pump}} = 2.5~\unit{\ps^{-1}\um^{-2}}$,
probe amplitude $S_{0,\text{probe}} = 0.5~\unit{\ps^{-1}\um^{-2}}$,
resonant frequency $\hbar \omega_\text{s} = 0.35~\unit{\meV}$ [see Methods];
$\kappa = 0.015~\unit{\meV\per\um}$. 
}
\label{fig:reconfig}
\end{figure*}

A prototypical system demonstrating a reconfigurable Chern interface in a polariton lattice is shown in Fig.~\ref{fig:reconfig}(a).
In particular, a non-resonant pump with amplitude below the condensate threshold [yellow shaded area] is used to populate an exciton reservoir that reaches a steady-state due to its inherent dissipation, while a resonant probe source [magenta star] excites the chiral edge modes.
As such, after the exciton reservoir becomes sufficiently populated, the polaritons experience a blueshift due to the repulsive nonlinear interactions according to Eqs.~\eqref{eq:polariton_rate}-\eqref{eq:reservoir_rate}, locally modifying the polariton's topology, and creating new topological interfaces that were not present in the unpumped system.
Once the non-resonant pump is turned off, the exciton reservoir's population dissipates, returning the polariton's local topology to its original configuration.

The complete temporal evolution of our non-resonantly pumped exciton-polariton lattice is shown in Figs.~\ref{fig:reconfig}(b)-(f), where we consider a system that is initially pumped in one region, but that pump is later turned off.
The dynamics of the local blueshift $E_\text{blueshift}(t) = g_r n_{\textrm{r}}(t)$ in the pumped region are shown in Fig.~\ref{fig:reconfig}(b), which demonstrates that the population of the exciton reservoir initially saturates at a steady-state (with $E_\text{blueshift} = 0.55~\unit{\meV}$) after a characteristic time given by $\gamma_r$ before becoming completely depleted once the pump amplitude is switched off.
Moreover, throughout the evolution of the reservoir population, real-space snapshots of the topological dynamics of this non-Hermitian and non-linear system can be found using the spectral localizer framework, as shown in the bottom panels of Figs.~\ref{fig:reconfig}(d)-(f).
In particular, given the tiny extent of the pumped region, topological band theory would only provide qualitative insights of the topology of the system, which may break down for this region's size. 
Instead, by monitoring the real parts of the spectral localizer's eigenvalues $\text{Re} \left[ \text{spec} \left( L_{(x,y_0,E_0)} \right) \right]$ for choices of $\vec{x}$ along the gray dashed line in Fig.~\ref{fig:reconfig}(a) and at the topological mode's expected energy $E_0 = 0.35~\unit{\meV}$ [see Fig.~\ref{fig:band}], the change in the polariton's topology can be directly observed in a quantitative and rigorous manner, as this spectral flow (at a given $t$) is responsible for shifts in $C_{(x,y_0,E_0)}^{\textrm{L}}$.
At the outer edge of the lattice, one of the eigenvalues of the spectral localizer crosses zero and the local Chern number changes from $C_{(x,y_0,E_0)}^{\textrm{L}} = 0 \rightarrow 1$ (as $x$ varies from outside the lattice to inside) because the lattice is topologically non-trivial while the surrounding empty space is trivial.
In the presence of the populated exciton reservoir, there is an additional change in the polariton's topology from the shape of the non-resonant pump: the local Chern number is trivial $C_{(\vec{x},E)}^{\textrm{L}} = 0$ inside the pumped region [see bottom panels of Fig.~\ref{fig:reconfig}(d)]. 
Once the pump is turned off [Fig.~\ref{fig:reconfig}(e)-(f)], the polariton's nonlinear interactions with the exciton reservoir dissipate and the blueshift eventually becomes negligible, such that the previously pumped region becomes topological again, removing the in-lattice topological boundary [see bottom panels of Fig.~\ref{fig:reconfig}(e)-(f)].

Overall, the picosecond time-scale reconfigurability of the exciton-polariton lattices topology can only be realized thanks to the driven-dissipative nature of the system, especially when the intrinsic dissipation $\gamma_r$ manifests in a different sector than the system' states to avoid its influence on the protected transport.
While an intrinsic loss $\gamma_c$ is present for the polariton state, this loss is only included to model realistic exciton-polariton system, and is by no means necessary to realize fast reconfigurable Chern topology.
Specifically, the dynamics of the excited polariton states, shown with the real-space snapshots in Figs.~\ref{fig:reconfig}(d)-(f), can be summarized as follows:
When the lattice is (locally) pumped, the excited polariton state at the bottom-left side of the lattice initially propagates along the edge of the topologically non-trivial lattice, before turning upward along the nonlinearly induced topological interface without being back-reflected [Fig.~\ref{fig:reconfig}(d)], as the pumped region is a trivial insulator at the excited polariton energy $E = 0.35~\unit{\meV}$.
However, when the non-resonant pump is turned off, the exciton reservoir dissipates and the previously pumped region returns to being topological. 
As such, there is no longer a topological interface in the system's bulk, and the topological edge mode that was previously propagating in the bulk starts to decay as the system's bulk becomes gapped again [Fig.~\ref{fig:reconfig}(e)].
Subsequently, the excited polariton propagates along the lattice edge without being back-reflected [Fig.~\ref{fig:reconfig}(f)].
The dynamics of the chiral edge mode's propagation can be quantified by looking at the intensity at different output locations, as shown in Fig.~\ref{fig:reconfig}(c).
The output intensity at the dark red cross $I_\text{out,1}$ reaches a plateau when the lattice is non-resonantly pumped, while the output intensity at the dark green cross $I_\text{out,2}$ reaches a plateau when the pump is turned off.
Notably, the green plateau (output $I_\text{out,2}$) exhibits a rapid switching time of approximately $70~\unit{\ps}$ and a high on/off state ratio, making it a promising candidate for ultrafast optical-switching~\cite{Chai2017}.
Altogether, with the base energy of the exciton-polariton being about $1~\unit{\eV}$~\cite{Karzig2015, Bardyn2015, Nalitov2015, Yi2016, Klembt2018, Solnyshkov2021} [see Methods], fast reconfigurable topologically protected routing at telecommunications wavelengths is achieved solely from the dynamical control over the shape and existence of a non-resonant pump applied to the non-trivial polariton lattice, which can be realized using spatial light modulators or electrical pumping~\cite{Schneider2013, Suchomel2018} [see also Sect.~\ref{sect_supp:exp_method} for further details~\cite{supp}].
%


\subsection{Multi-channel topological routing with multiple non-resonant pump patterns}
\label{sect:multi_channel}

\begin{figure}[!]
\center
\includegraphics[width=\columnwidth]{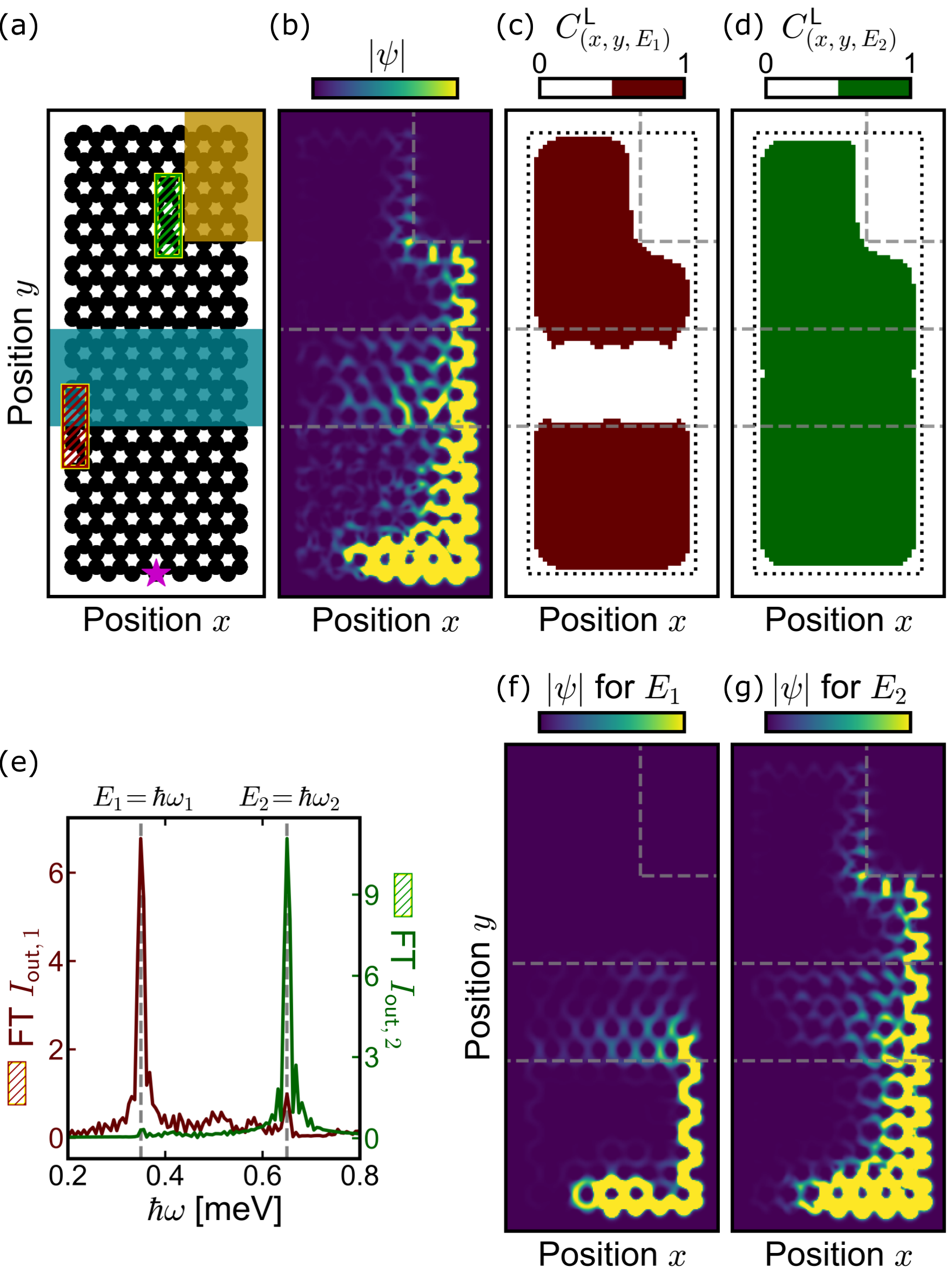}
\caption{
\textbf{Frequency-dependent reconfigurable topological routing with non-resonant pumpings.}
(a) Potential landscape $V(x,y)$ of the polariton lattice arrange in a honeycomb lattice.
The teal (yellow) shaded area depicts the pump pattern with pump amplitude $S_{0,\text{pump}}^{(0)}$ ($S_{0,\text{pump}}^{(1)}$), the magenta star indicates the position of the probe source.
(b) Snapshot of the total intensity of the polariton $|\psi|^2 = |\psi_+|^2 + |\psi_-|^2$.
(c),(d) Local Chern numbers $C_{(x,y,E_i)}^{\textrm{L}}$ calculated over the system at the energy $E_i = E_0 = 0.35~\unit{\meV}$ and $E_i = E_0 = 0.65~\unit{\meV}$, respectively.
(e) Fourier transform (FT) of the polariton signal over the red and green hatched areas in (a).
(f),(g) Snapshot of the total intensity of the polariton $|\psi|^2$ when the probe source is resonantly exciting only either at the $E_1 = \hbar \omega_{\text{s},1}$ or $E_2 = \hbar \omega_{\text{s},2}$ energy, respectively.
In panels (b)-(d),(f)-(g), the snapshot of the topological dynamics has been taken at $t = 658~\unit{\ps}$, and the gray dashed lines delimit the non-resonant pump patterns.
The black dotted lines in (c),(d) delimit the lattice.
The parameter values for the Hamiltonian are the same as in Fig.~\ref{fig:band}.
Dynamical parameter values:
pump amplitude for the teal area $S_{0,\text{pump}} = S_{0,\text{pump}}^{(0)} = 0.5~\unit{\ps^{-1}\um^{-2}}$,
pump amplitude for the orang area $S_{0,\text{pump}} = S_{0,\text{pump}}^{(1)} = 2.7~\unit{\ps^{-1}\um^{-2}}$,
probe amplitude $S_{0,\text{probe}} = 0.5~\unit{\ps^{-1}\um^{-2}}$ 
with resonant frequencies $\hbar \omega_{\text{s},1} = 0.35~\unit{\meV}$ and $\hbar \omega_{\text{s},2} = 0.65~\unit{\meV}$ [see Methods];
$\kappa = 0.015~\unit{\meV\per\um}$. 
}
\label{fig:multi_channel}
\end{figure}

Beyond reconfigurable single-channel routing demonstrated in Sect.~\ref{sect:reconfig} and Fig.~\ref{fig:reconfig}, the use of multiple pump patterns in the system enables reconfigurable multi-channel topological routing. 
While a straightforward extension of the previous results provides a recipe to simultaneously guide information along topological interface modes from different input channels to different output channels at a single frequency, the use of multiple pump patterns with different amplitudes also enables frequency-dependent reconfigurable topological routing where the propagation path of the topological edge states varies based on their frequency.

To illustrate reconfigurable multi-channel topological routing, we consider a prototypical exciton-polariton lattice where two non-resonant pump patterns with different pump amplitudes are utilized, as shown in Fig.~\ref{fig:multi_channel}, while a single probe source is used to excites topological edge modes at several energies.
In particular, one pump pattern has a pump amplitude small enough to only induce a change of the system topology in a lower energy range, while the second pump is strong enough to modify the system's topology in a higher energy range.
Altogether, the different pump amplitudes induce local blueshifts of different strengths, resulting in energy-dependent modifications to the system's local topology and thus nonlinearly induced topological interfaces that are distinct for inputs at different energies.  

A real-space snapshot of the exciton-polariton lattice's topological dynamics is shown in Fig.~\ref{fig:multi_channel}(b)-(d), where the nonlinearly induced blueshifts of the non-resonant pump in the teal and yellow shaded areas are respectively $E_\text{blueshift} = 0.11~\unit{\meV}$ and $E_\text{blueshift} = 0.6~\unit{\meV}$ once the corresponding population of the exciton reservoir reached a steady state.
As such, the local blueshift in the teal shaded area leads to a gapless ``heterostructure'' interface configuration which is topological for the lower band gap energy range while being trivial for the higher band gap energy range [see Figures~\ref{fig_supp:band_gapless}-\ref{fig_supp:reconfig_gapless} in Sect.~\ref{sect_supp:band_gapless} for further details~\cite{supp}].
In contrast, the local blueshift in the yellow shaded area is similar to Fig.~\ref{fig:band} and Fig.~\ref{fig:reconfig} where there is a nonlinearly induced topological interface spanning the second band gap energy range.
Thus, the topologically distinct domains of the system at energies in the lower and higher energy ranges are different, as plotted in Fig.~\ref{fig:multi_channel}(c)-(d). 
Overall, an excitation at $E_1$ in the lower energy range propagates along the topological interface induced by the teal non-resonant pump [Fig.~\ref{fig:multi_channel}(f)], while an excitation at $E_2$ in the higher energy range propagates through the teal pump area while still experiencing the topological interface formed by the yellow shaded area [Fig.~\ref{fig:multi_channel}(g)].
Therefore, for a point source resonantly exciting topological modes at energies $E_1$ and $E_2$, the two topological modes will be guided to different output channels [Fig.~\ref{fig:multi_channel}(b)]; taking the Fourier transform (FT) of the polariton time-evolution summed over the different output-channel areas [see Fig.~\ref{fig:multi_channel}(a)] leads to the respective peaks at energies $E_1$ and $E_2$ [Fig.~\ref{fig:multi_channel}(d)]. Altogether, our results demonstrate multi-channel (frequency-dependent) topological routing that can be dynamically controlled by multiple non-resonant pumps.


\section{Discussion}

To summarize, we have presented a technologically realistic proposal for dynamic reconfigurable topological routing in nonlinear driven-dissipative photonic systems in a single device, and shown that this approach works at telecommunication wavelengths and at picosecond time scales.
This method has been illustrated with non-resonantly pumped exciton-polariton lattices, and in particular, using a continuum model of the driven-dissipative Gross-Pitaevskii equations parameterized with prior experimental results~\cite{Klembt2018}.
Moreover, we have demonstrated dynamic control over the propagation path of the system's chiral edge states -- including its frequency-dependent behavior -- through a nonlinearly induced topological phase transition stemming from spatially non-uniform non-resonant pumping of the exciton reservoir.
We resolved the system's local topological phase change by generalizing the spectral localizer framework to accommodate nonlinear non-Hermitian systems, and find a broad range of pump powers and polarizations are suitable for inducing a local nonlinear topological transition in exciton-polariton lattices~\cite{supp}.
More broadly, our framework enables a rigorous and quantitative probing of the dynamical change of the topology over time and within regions containing very few unit cells where topological band theory would be unreliable.
As such, we anticipate the spectral localizer to prove valuable for characterizing topological mechanisms unique to nonlinear, non-Hermitian systems.

Overall, compared with previous proposals operating at the telecommunication regime~\cite{Dai2024, Zhao2019}, our approach represents a significant technological advancement, avoiding large losses that hinder topological transport and overcoming the limitations of unscalable multidimensional dynamic control over each element in a system, and uniquely enables reconfigurable multi-channel topological routing, as opposed to single channel routing.
Additionally, while reconfigurable topology has been realized one-dimensional (1D) systems~\cite{Pieczarka2021, Mandal2023, Lackner2024, Gagel2022}, these 1D topological systems only protect stationary states, and fundamentally differ from the propagating edge states in 2D Chern insulators.
Looking forward, the possibility of realizing reconfigurable topological Chern photonic devices provides a new route to achieving reflectionless non-reciprocal routing, a capability with applications in a variety of communication technologies.
Furthermore, as dynamical tuning of a system's local topology is rooted in external control of a system's nonlinear interactions, we expect nonlinearly induced topological phase transitions can be achieved in a variety of other pump-probe configurations, such as the Stark effect via electrical gating~\cite{Hayat2012, Cancellieri2014, Panna2019, Gagel2022}, in platforms featuring multiwave mixing~\cite{Pilozzi2017, Jia2023, Mittal2021, Zhang2019} through a second illumination source or perturbation, or in other strongly coupled composite systems such as phonons-polariton lattices with induced phonon-phonon interactions~\cite{Ravichandran2020}.


\section{Methods}

\subsection{Numerical methods and parameter values}
\label{sect:methods}

The polariton lattice is a honeycomb lattice with lattice constant $a = 2.95~\unit{\um}$ (center-to-center distance is $1.7~\unit{\um}$), 
and rod radius $1~\unit{\um}$.
The parameters of the Hamiltonian are chosen based on Ref.~\cite{Klembt2018} where
the polariton mass is $m = 1.3 \times 10^{-4} m_0$ with $m_0$ the free electron mass, 
the spin-orbit coupling strength is $\beta_\text{eff} = 0.2~\unit{\meV.\um^2}$ and
the Zeeman coefficient is $\Delta_\text{eff} = -0.3~\unit{\meV}$.
Without loss of generality, the bare energy of the exciton-polaritons $\tilde{E_0} \sim 1~\unit{\eV}$~\cite{Karzig2015, Bardyn2015, Nalitov2015, Yi2016, Klembt2018, Solnyshkov2021} has not been taken into account in the Hamiltonian, as $\tilde{E_0}$ only accounts for an energy shift of all the bands.
The ribbon band structure is calculated from the ribbon geometry with periodic boundary conditions on the left and right boundaries, and Dirichlet boundary conditions on the top and bottom boundaries.

To integrate the rate equations given by the driven-dissipative Gross-Pitaevskii equations [Eqs.~\eqref{eq:polariton_rate}-\eqref{eq:reservoir_rate}], the time $t$ is re-scaled to $t'=t/\hbar$ and a third-order Adams-Bashforth method is utilized with a time step $dt' = 5 \times 10^{-3}$ and grid mesh of $dx = dy = 0.151~\unit{\um}$.
The dynamical parameters are such that 
the relaxation rates for the polariton state and exciton reservoir are $\gamma_c = 0.03~\unit{\per\ps}$ and $\gamma_r = 1.5 \gamma_c$, 
the polariton-polariton and polariton-exciton interaction strengths are $g_c = 5 \times 10^{-3}~\unit{\meV.\um^2}$ and $g_r = 10 \times 10^{-3}~\unit{\meV.\um^2}$,
the amplification rate of the polariton state due to stimulated scattering of polariton from the reservoir is $R = 3 \times 10^{-4}~\unit{\ps^{-1}\um^2}$.

In Figure~\ref{fig:reconfig}, the pump is given by a top-hat-like function of the time-interval $t_0 = 10 [\hbar]$ and $t_1 = 850 [\hbar]$ [see Fig.~\ref{fig:reconfig}(b)]
\begin{equation}
S_{\text{pump}}
= \frac{1}{1 + e^{-\frac{t-t_0}{2\tau}}} \left( 1 - \frac{1}{1 + e^{-\frac{t-t_1}{2\tau}}} \right)
\left(
\begin{array}{c}
S_{0,\text{pump},+} \mathbf{1} \\[1.ex] 
S_{0,\text{pump},-} \mathbf{1}
\end{array}
\right)
,
\end{equation}
with $\tau_t = 0.8 [\hbar]$ and
pump amplitude $S_{0,\text{pump},\pm} = S_{0,\text{pump}} = 2.5~\unit{\ps^{-1}\um^{-2}}$.
The resonant probe is a continuous wave source, centered at $(x_\text{s}, y_\text{s})$ [see magenta star in Fig. a] 
\begin{equation}
S_{\text{probe}}
= e^{ -\frac{(x_\text{s}-x)^2+(y_\text{s}-y)^2}{2\tau_{xy}^2} } e^{-i \omega_\text{s} t} e^{i k_\text{s} x}
\left(
\begin{array}{c}
S_{0,\text{probe},+} \mathbf{1} \\[1.ex] 
S_{0,\text{probe},-} \mathbf{1}
\end{array}
\right)
,
\end{equation}
with $\tau_{xy} = 1~\unit{\um}$, 
probe amplitude $S_{0,\text{probe},\pm} = S_{0,\text{probe}} = 0.5~\unit{\ps^{-1}\um^{-2}}$,
resonant frequency $\hbar \omega_\text{s} = 0.35~\unit{\meV}$, and
resonant wavevector $k_\text{s} = 0.6 [\pi/a]$.

In Figure~\ref{fig:multi_channel}, the probe takes the form 
\begin{align}
\begin{split}
S_{\text{probe}}
& = e^{ -\frac{(x_\text{s}-x)^2+(y_\text{s}-y)^2}{2\tau_{xy}^2} } 
\left(
\begin{array}{c}
S_{0,\text{probe},+} \mathbf{1} \\[1.ex] 
S_{0,\text{probe},-} \mathbf{1}
\end{array}
\right)
\\
& \qquad \qquad \times
\left( e^{-i \omega_{\text{s},1} t} e^{i k_{\text{s},q} x} + e^{-i \omega_{\text{s},2} t} e^{i k_{\text{s},2} x} \right)
,
\end{split}
\end{align}
with $\tau_{xy} = 1~\unit{\um}$, 
probe amplitude $S_{0,\text{probe},\pm} = S_{0,\text{probe}} = 0.5~\unit{\ps^{-1}\um^{-2}}$,
resonant frequencies $\hbar \omega_{\text{s},1} = 0.35~\unit{\meV}$ and $\hbar \omega_{\text{s},2} = 0.65~\unit{\meV}$, and
resonant wavevectors $k_{\text{s},1} = 0.62 [\pi/a]$ and $k_{\text{s},2} = 0.47 [\pi/a]$.


\section*{Acknowledgments}

\textbf{Funding:} 
S.W.\ acknowledges support from the Laboratory Directed Research and Development program at Sandia National Laboratories.
S.B.\ and S.H.\ acknowledge financial support by the German Research Foundation (DFG) under Germany’s Excellence Strategy–EXC2147 “ct.qmat” (project id 390858490) and HO 5194/12-1.
A.C.\ acknowledges support from the U.S.\ Department of Energy, Office of Basic Energy Sciences, Division of Materials Sciences and Engineering.
This work was performed in part at the Center for Integrated Nanotechnologies, an Office of Science User Facility operated for the U.S. Department of Energy (DOE) Office of Science.
Sandia National Laboratories is a multimission laboratory managed and operated by National Technology \& Engineering Solutions of Sandia, LLC, a wholly owned subsidiary of Honeywell International, Inc., for the U.S. DOE's National Nuclear Security Administration under Contract No. DE-NA-0003525. 
The views expressed in the article do not necessarily represent the views of the U.S. DOE or the United States Government.

\bibliography{ref}



\newpage

\renewcommand{\thesection}{Supplementary note \arabic{section}}
\setcounter{section}{0}

\renewcommand{\thefigure}{S\arabic{figure}}
\setcounter{figure}{0}

\renewcommand{\theequation}{S\arabic{equation}}
\setcounter{equation}{0}



\section{Probing the topology in polariton lattices with blueshift}
\label{sect_supp:localizer}

\begin{figure}[t!]
\center
\includegraphics[width=\columnwidth]{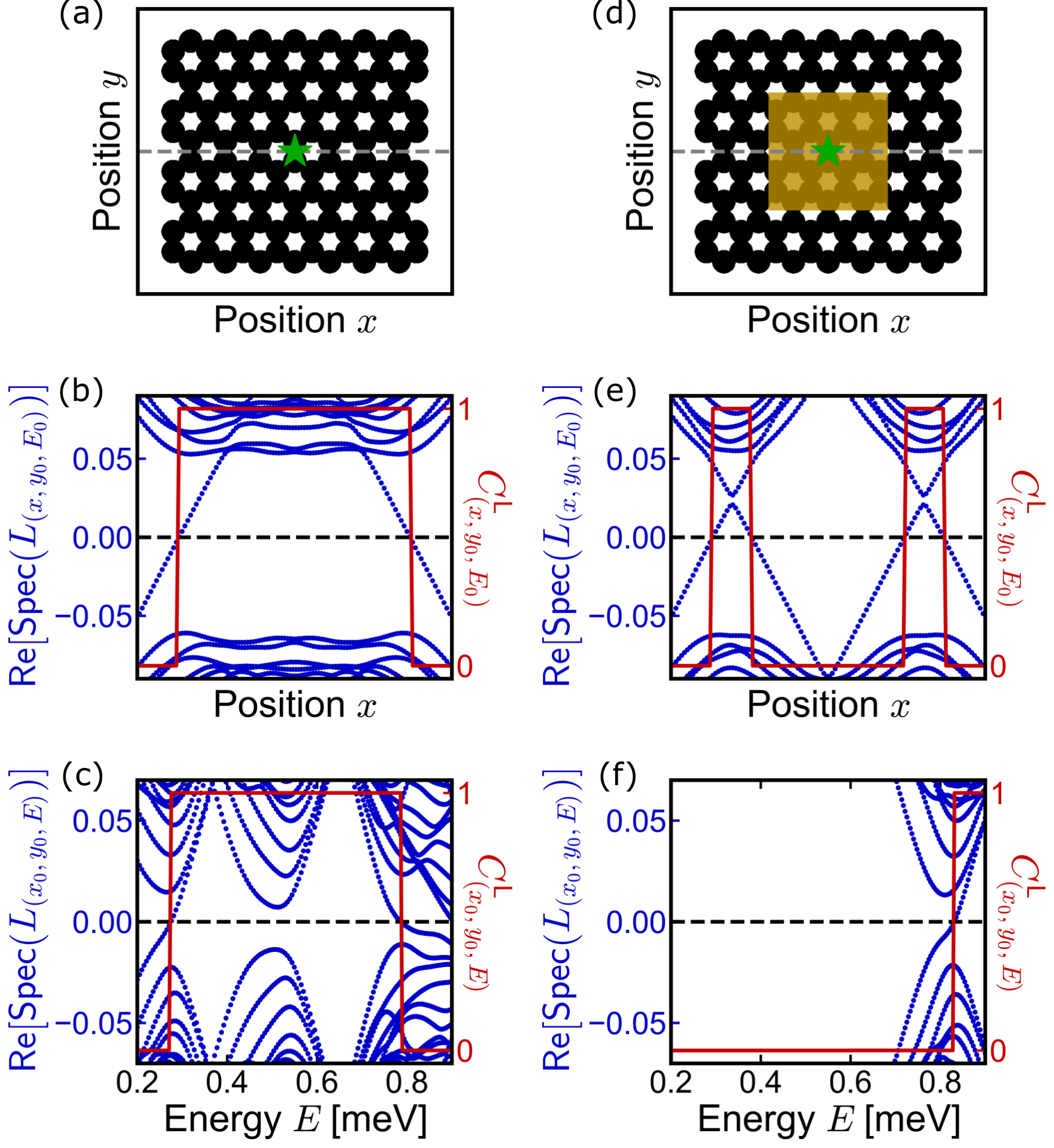}
\caption{
\textbf{Probing topology of non-resonantly pumped systems using the spectral localizer framework.}
(a) Potential landscape $V(\vec{x})$ of the polariton system arranged in a honeycomb lattice. 
The black and white regions correspond to potentials of $V=6~\unit{\meV}$ and $V=0~\unit{\meV}$, respectively.
(b) Eigenvalues of the spectral localizer $\text{Re} \left[ \text{Spec} \left( L_{(x,y_0,E_0)} \right) \right]$ and the local Chern number $C_{(x,y_0,E_0)}^{\textrm{L}}$ along the dashed line in (a) and at $E_0=0.35~\unit{\meV}$.
(c) Eigenvalues of the spectral localizer $\text{Re} \left[ \text{Spec} \left( L_{(x_0,y_0,E)} \right) \right]$ and the local Chern number $C_{(x_0,y_0,E)}^{\textrm{L}}$ along the energy axis and at the position of green star in (a).
(d)-(f) Same as (a)-(c) but with an yellow overlay on the lattice in (d), depicting the blueshift.
Parameter values:
lattice constant is $a = 2.95~\unit{\um}$ (center-to-center distance is $1.7~\unit{\um}$), 
radius of the rods is $1~\unit{\um}$;
$m = 1.3 \times 10^{-4} m_0$ with $m_0$ the free electron mass, 
$\beta_\text{eff} = 0.2~\unit{\meV.\um^2}$,
$\Delta_\text{eff} = -0.3~\unit{\meV}$;
blueshift energy used is $E_\text{blueshift}=0.55~\unit{\meV}$;
$\kappa = 0.015~\unit{\meV\per\um}$. 
}
\label{fig_supp:localizer}
\end{figure}

In this section, we present how the system's local topology has been determined for a system with no blueshift and one that is partly blueshifted [see for example Figs.~2(b),(e),(h) and Figs.~2(d)-(f) in the main text].
The topology in the polariton lattice is characterized directly from the continuous model with the spectral localizer framework.
In particular, the topology is probed by looking at the real parts of the spectral localizer's eigenvalues denoted $\text{Re} \left[ \text{Spec} \left( L_{(x,y,E)} \right) \right]$, and the corresponding local Chern number $C_{(x,y,E)}^{\textrm{L}}$.

In the absence of any nonlinearly induced blueshift [Fig.~\ref{fig_supp:localizer}(a)], the lattice is topologically non-trivial.
A real-space picture of the topology is shown in Figure~\ref{fig_supp:localizer}(b) by plotting the spectral flow of the localizer $\text{Re} \left[ \text{Spec} \left( L_{(x,y_0,E_0)} \right) \right]$ local Chern number $C_{(x,y_0,E_0)}^{\textrm{L}}$ when varying $x$ with fixed $y_0$ along the gray dashed line in Fig.~\ref{fig_supp:localizer}, and at $E_0 = 0.35~\unit{\meV}$.
The eigenvalues of the localizer cross the zero axis around the edges of the lattice, indicating a change of local Chern number and the non-trivial topology of the lattice.
Similarly, the topology can also be spectrally resolved by varying $(x_0,y_0,E)$ along the energy axis with a fixed position $(x_0, y_0)$.
Figure~\ref{fig_supp:localizer}(c) displays $\text{Re} \left[ \text{Spec} \left( L_{(x_0,y_0,E)} \right) \right]$ and $C_{(x_0,y_0,E)}^{\textrm{L}}$, for a fixed spatial position $(x_0, y_0)$ given by the green start in Fig.~\ref{fig_supp:localizer}(a).

With a finite partial blueshift of the lattice [Fig.~\ref{fig_supp:localizer}(d)], $B = E_\text{blueshift} I$ and $\Gamma = 0$, the topology is only changed locally.
By looking at the spectral localizer's spectrum along the dashed line, we can see a change of the topology inside the blueshifted region: There is additional crossing of zero for $\text{Re} \left[ \text{Spec} \left( L_{(x,y_0,E_0)} \right) \right]$ and the local Chern number becomes trivial inside of the pumped region, as shown in Fig.~\ref{fig_supp:localizer}(e).
Moreover, the topology along the energy axis is also modified inside the blueshifted region.
The local Chern number is now trivial up to approximately $E=0.8~\unit{\meV}$ as shown in Fig.~\ref{fig_supp:localizer}(f), while there was non-trivial local Chern number for $E = 0.3-0.8~\unit{\meV}$ in the unpumped system [see Fig.~\ref{fig_supp:localizer}(c)].

The local Chern numbers shown in Fig.~1 in the main text correspond to the calculated local Chern number along the energy axis in Figs.~\ref{fig_supp:localizer}(c),(f).


\section{Reconfigurable topology robust to disorder}
\label{sect_supp:pert_localizer}

\begin{figure}[t]
\center
\includegraphics[width=\columnwidth]{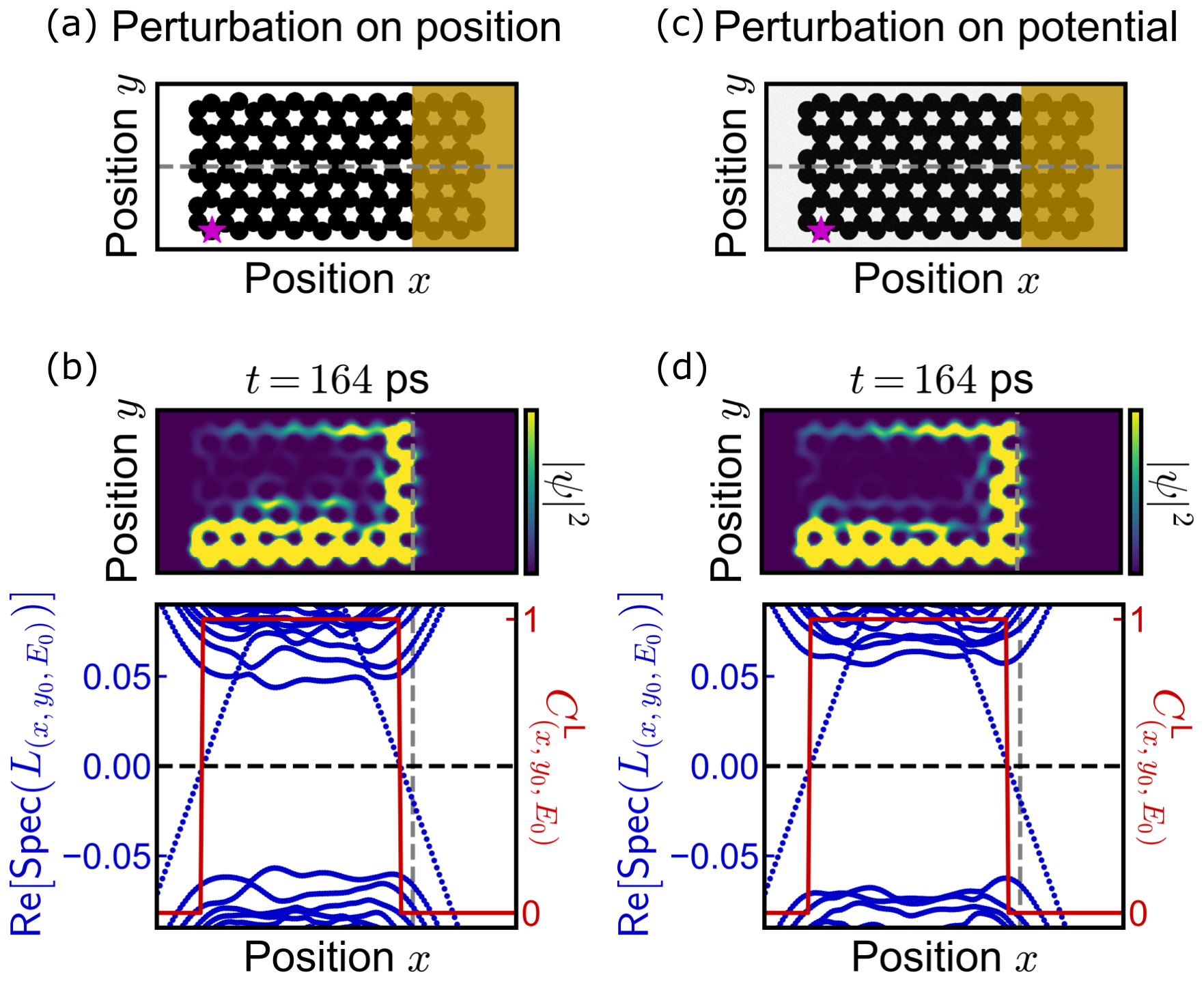}
\caption{
\textbf{Topological robustness for nonlinear-induced topological interfaces.}
(a) Potential landscape $V(\vec{x})$ of the polariton system arranged in a honeycomb lattice, with $10\%$ perturbation on the quantum wells' position. 
The black and white regions correspond to potentials of $V=6~\unit{\meV}$ and $V=0~\unit{\meV}$, respectively.
The yellow shaded area depicts the pump pattern, the magenta star indicates the position of the probe source.
(b) Snapshot of the total intensity of the polariton $|\psi|^2 = |\psi_+|^2 + |\psi_-|^2$, with the corresponding eigenvalues of the spectral localizer $\text{Re} \left[ \text{Spec} \left( L_{(x,y_0,E_0)} \right) \right]$ and the local Chern number $C_{(x,y_0,E_0)}^{\textrm{L}}$ along the gray dashed line in (a) and at $E_0 = 0.35~\unit{\meV}$.
(c)-(d) Same as (a)-(b) but with perturbation on the polariton potential of strength $0.3~\unit{\meV}$.
The parameter values for the Hamiltonian, and the dynamics are the same as in Fig.~\ref{fig:reconfig} in the main text.
}
\label{fig_supp:pert_localizer}
\end{figure}

This section demonstrates that the newly formed topological interfaces are robust against disorder.
In particular, disorder in the position of the quantum wells and in the polariton potential are considered.

For example, for a $10\%$ disorder in position (drawn from a uniform distribution), the dynamics of the topological edge states and the system's local Chern topology is plotted in Fig.~\ref{fig_supp:pert_localizer}(a)-(b), thus showing the robustness of the newly-formed topological interface (and of the topological edge states) against shift in position of the quantum wells.
Similarly, one can show in Fig.~\ref{fig_supp:pert_localizer}(c)-(d) that newly-formed topological interface and of the topological edge states are also robust against perturbation in the polariton potential, with perturbation strength of $0.3~\unit{\meV}$ drawn from a uniform distribution.


\section{Reconfigurable topological routing with small blueshift}
\label{sect_supp:band_gapless}

\begin{figure}[!]
\center
\includegraphics[width=\columnwidth]{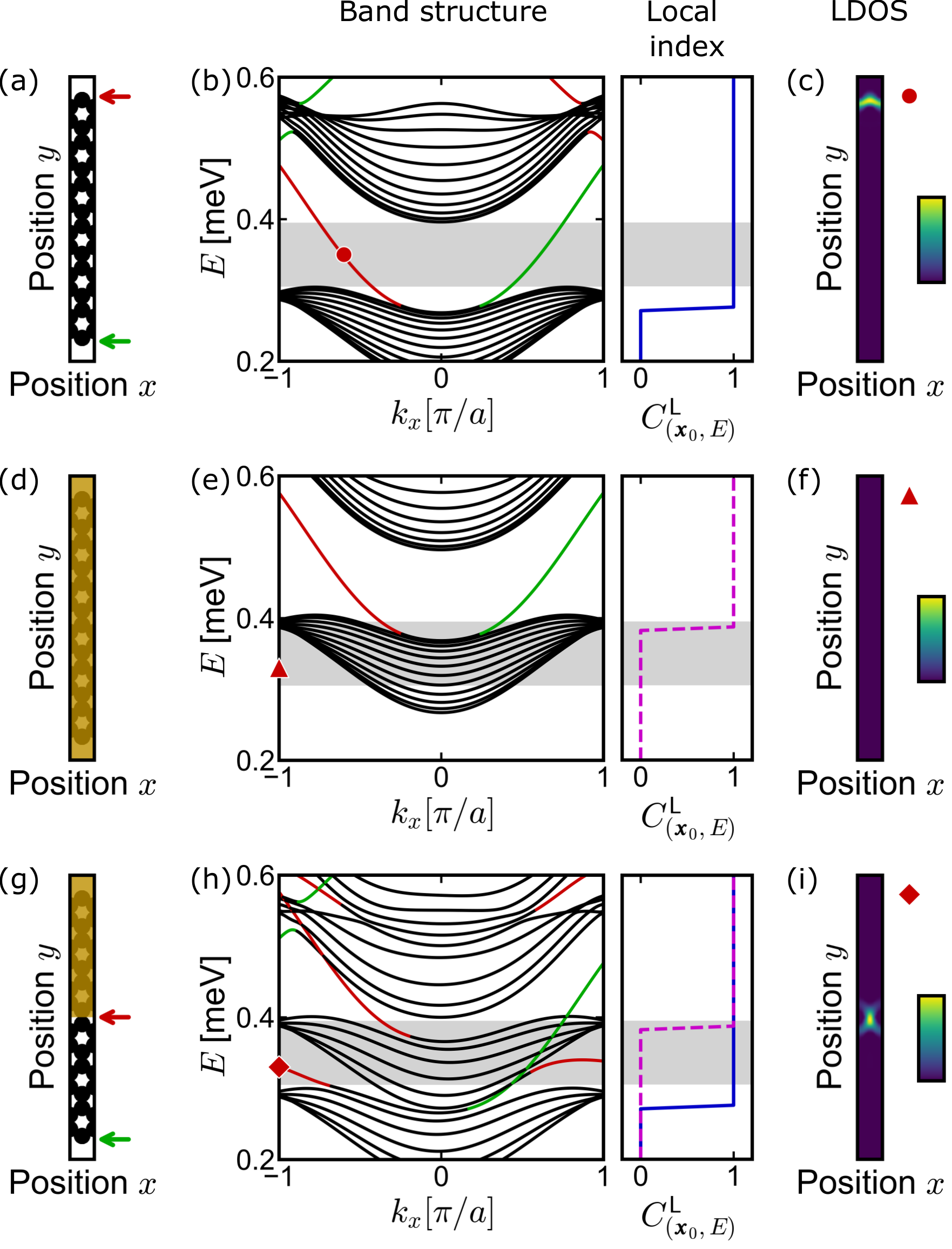}
\caption{
\textbf{Change of the topology due to the blueshift, gapless case.}
(a) Potential landscape $V(\vec{x})$ of the ribbon polariton lattice arranged in a honeycomb lattice. 
The black and white region correspond to a potential of $V=6~\unit{\meV}$ and $V=0~\unit{\meV}$, respectively.
(b) Ribbon band structure of (a), and the corresponding local Chern number $C_{(\vec{x}_0,E)}^{\textrm{L}}$ , at a the given energy, calculated using the spectral localizer.
In the band structure, the black lines correspond to the bulk modes, and the green (red) lines denotes the topological edge mode dispersion localized at the bottom (top) side of the lattice, as shown by the color-coded arrows in (a).
The gray shaded area indicates the first band gap, which is the energy range of interest. 
(c) Local density of states (LDOS) of the red line at $E=0.35~\unit{\meV}$, $k_x = -0.6 [\pi/a]$. 
(d)-(f) Same as (a)-(c) but with an yellow overlay depicting the blueshift of the whole ribbon structure.
In (f), the LDOS is plotted for $E=0.33~\unit{\meV}$, $k_x = -1 [\pi/a]$.
(g)-(i) Same as (a)-(c) but with the blueshift applied only to half of the ribbon lattice. 
The red (and green) lines in the band structure correspond to the topological edge modes localized at the interface between the blueshifted and non-bleushifted areas (and at the bottom edge of the lattice). 
In (h), the solid blue line and dashed magenta line are the local Chern number calculated inside the non-blueshifted and blueshifted regions, respectively.
(i) LDOS of the red line at $E=0.33~\unit{\meV}$, $k_x = -1 [\pi/a]$. 
Parameter values:
The lattice constant is $a = 2.95~\unit{\um}$ (center-to-center is $1.7~\unit{\um}$), 
radius of the rods is $1~\unit{\um}$;
$m = 1.3 \times 10^{-4} m_0$ with $m_0$ the free electron mass, 
$\beta_\text{eff} = 0.2~\unit{\meV.\um^2}$,
$\Delta_\text{eff} = -0.3~\unit{\meV}$;
blueshift energy used is $E_\text{blueshift}=0.1~\unit{\meV}$;
$\kappa = 0.02~\unit{\meV\per\um}$. 
}
\label{fig_supp:band_gapless}
\end{figure}

Here, we show that the proposed method for reconfigurable topological routing also works if the blueshifted region becomes gapless in the energy range of interest.
In particular, we discuss the case of a small blueshift, leading to an interface between a topological non-trivial gapped region with a topologically trivial system that is gapless in the same energy range.

In the absence of blueshift, the lattice is topologically non-trivial and its band structure and topology are similar to Figs.~2(a)-(c) in the main text [see Figs.~\ref{fig_supp:band_gapless}(a)-(c)].
However, with a small blueshift of $E_\text{blueshift} = 0.1~\unit{\meV}$, $\Gamma = 0$, depicted by the yellow overlay in Fig.~\ref{fig_supp:band_gapless}(d), the bands are only slightly blueshifted, resulting in the bulk bands spectrally overlapping with the energy of interest [see gray shaded area in Fig.~\ref{fig_supp:band_gapless}(e)].
Thus, this partially blueshifted system is gapless.
Nevertheless, even though the system is gapless in the relevant energy range, the topology can still be classified using the spectral localizer~\cite{Dixon2023, Wong2024}.
The right panel of Fig.~\ref{fig_supp:band_gapless}(e) shows the local Chern number, calculated using similar methods as in Sect.~\ref{sect_supp:localizer}, demonstrating that the blueshifted portion of the system is topologically trivial in part of the gray shaded energy range.
Therefore, a bulk-edge correspondence is possible at the interface between the blueshifted and non-blueshifted regions. 
Although the system is gapless, namely there is no complete shared band gap in the nonlinear heterostructure, the system features an incomplete band gap where the topological mode appears [Fig.~\ref{fig_supp:band_gapless}(h)] due to the change of the local topology across the interface.
The red (and green) lines correspond to the topological mode localized at the newly formed topological interface (and bottom edge of the lattice), as shown through the LDOS of the red line at $E = 0.33~\unit{\meV}$ plotted in Fig.~\ref{fig_supp:band_gapless}(i).

\begin{figure}[!]
\center
\includegraphics[width=\columnwidth]{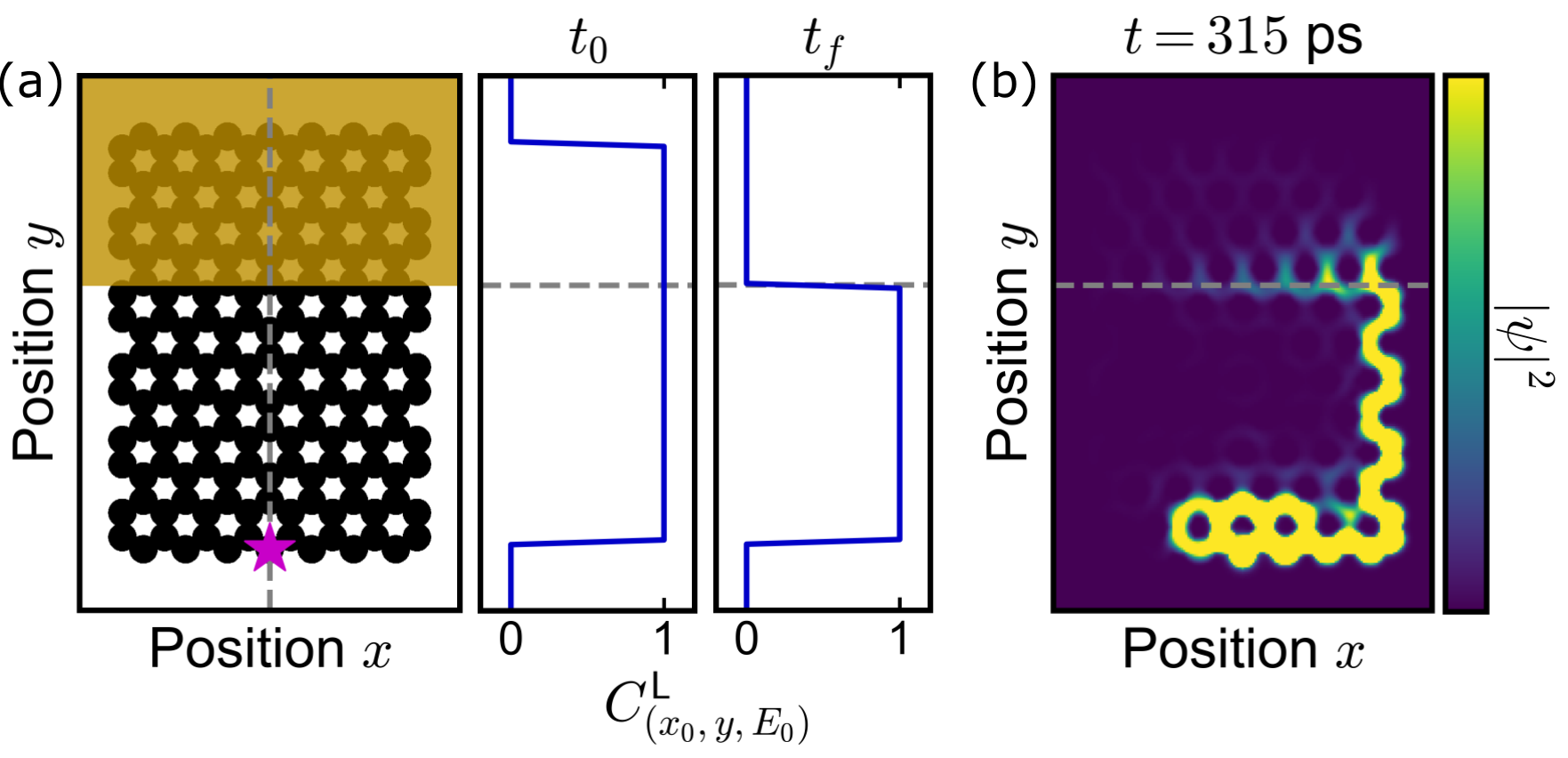}
\caption{
\textbf{Reconfigurable topological routing with non-resonant pumping, gapless case.}
(a) Potential landscape $V(\vec{x})$ of the polariton lattice arrange in a honeycomb lattice, and the local Chern number $C_{(\vec{x},E)}^{\textrm{L}}$ along the $x_0$ gray dashed line and at $E=0.33~\unit{\meV}$, calculated using the spectral localizer before the pump starts ($t_i$) and at the final time ($t_f$).
The yellow shaded area depicts the pump pattern, the magenta star indicates the position of the probe source.
(b) Snapshot of the total intensity of the polariton $|\psi|^2 = |\psi_+|^2 + |\psi_-|^2$.
The parameter values for the Hamiltonian are the same as in Fig.~\ref{fig_supp:band_gapless}.
Dynamical parameter values:
$\gamma_c = 0.03~\unit{\per\ps}$, 
$\gamma_r = 1.5 \gamma_c$, 
$g_c = 5 \times 10^{-3}~\unit{\meV.\um^2}$,
$g_r = 10 \times 10^{-3}~\unit{\meV.\um^2}$,
$R = 3 \times 10^{-4}~\unit{ps^{-1}\um^2}$;
$S_{0,\text{pump}} = 0.5~\unit{\ps^{-1}\um^{-2}}$,
$S_{0,\text{probe}} = 0.5~\unit{\ps^{-1}\um^{-2}}$;
$\hbar \omega_\text{s} = 0.35~\unit{\meV}$,
$k_\text{s} = 1 [\pi/a]$;
$\kappa = 0.015~\unit{\meV\per\um}$.
}
\label{fig_supp:reconfig_gapless}
\end{figure}

The dynamic behavior of the nonlinearly induced gapless topological heterostructure is shown in Fig.~\ref{fig_supp:reconfig_gapless}.
The calculation of the local Chern number along the gray dashed line [Fig.~\ref{fig_supp:reconfig_gapless}(a)] illustrates the change of topology at $E_0 = 0.33~\unit{\meV}$ inside gapless energy range.
The system becomes trivial inside the pumped region, as calculated at the final time ($t_f$).
Consequently, using a resonant probe source with a frequency and wavevector chosen in the incomplete band gap (see caption of Fig.~\ref{fig_supp:reconfig_gapless}), a topological resonance can be excited that propagates along the newly topological interface, as shown in Fig.~\ref{fig_supp:reconfig_gapless}(b). 
Here, the gapless nature of the trivial region means that the propagating edge mode is a resonance, and has some overlap with the available bulk states yielding another source of loss for this channel; indeed, some diffraction into the bulk can be seen in Fig.~\ref{fig_supp:reconfig_gapless}(b).


\section{Experimental realization scheme}
\label{sect_supp:exp_method}

The envisioned experiment would employ a two-dimensional lattice of coupled micropillar cavities, fabricated from a GaAs-based microcavity structure with embedded quantum wells, similar to those used in previous demonstrations of polaritonic topological insulators~\cite{Klembt2018}. 
The lattice geometry, typically honeycomb, would be chosen to support topological phases under appropriate symmetry-breaking perturbations, such as TE-TM splitting, magnetic fields, or staggered sublattice potentials.

Micropillars would have diameters in the range of $2.0-3.0~\unit{\um}$, with center-to-center distances of approximately $1.6-2.7~\unit{\um}$, depending on the pillar size, to ensure sufficient coupling between adjacent sites. 
The system would be excited non-resonantly using a continuous-wave laser beam tuned above the exciton resonance (e.g., $~740~\unit{\nm}$), thereby creating an incoherent exciton reservoir. 
The spatial profile of the pump would be structured using a spatial light modulator (SLM), enabling site-selective and dynamically reconfigurable injection of the exciton population. 
This reservoir modifies the polariton effective potential locally via repulsive interactions, allowing control over onsite energies and, consequently, the topological phase of different regions within the lattice.

The pump profile would be engineered to induce sharp boundaries between topologically trivial and nontrivial regions within a single, continuous lattice. 
The SLM’s spatial modulation resolution would be matched to the lattice scale using appropriate optics. 
Real-time dynamic reconfiguration of the domain wall could be achieved by updating the SLM pattern.

The edge state propagation would be probed using near-field photoluminescence imaging, collected through a microscope objective and projected onto a CCD camera. Momentum-space and energy-resolved measurements could be performed via Fourier-space imaging and spectral filtering, respectively. 
The time-resolved measurements using a streak camera would allow observation of edge state dynamics during topological transitions induced by time-varying pump profiles.

The topological characterization would rely on numerical evaluation of the spectral localizer from the measured or simulated system Hamiltonian, enabling a local, real-space determination of topological invariants, even in the presence of dissipation and inhomogeneous pumping. 
This framework is particularly well-suited to non-Hermitian, driven-dissipative systems and can be directly applied to validate the formation and manipulation of topological boundaries in the experiment.

This experimental design enables the observation of dynamically reconfigurable topological states in a photonic platform without requiring material modification or slow electro-optic control, relying instead on fast, reprogrammable optical pumping patterns to manipulate topological phases in real time.


\section{Topological routing with circular polarized pump pattern}

So far, we have considered polariton lattices that are topologically non-trivial at the energy range of interest in the absence of a nonlinearly induced blueshift.
However, with a gapless topologically trivial polariton lattice, namely without any external magnetic field $\Delta_\text{eff} = 0~\unit{\meV}$, the existence of topological modes and the dynamical control of their propagation path can also be realized using non-resonant circularly polarized pumps.
%


\subsection{Non-trivial topology induced by spin-dependent blueshift}
\label{sect_supp:band_spin}

\begin{figure}[t]
\center
\includegraphics[width=\columnwidth]{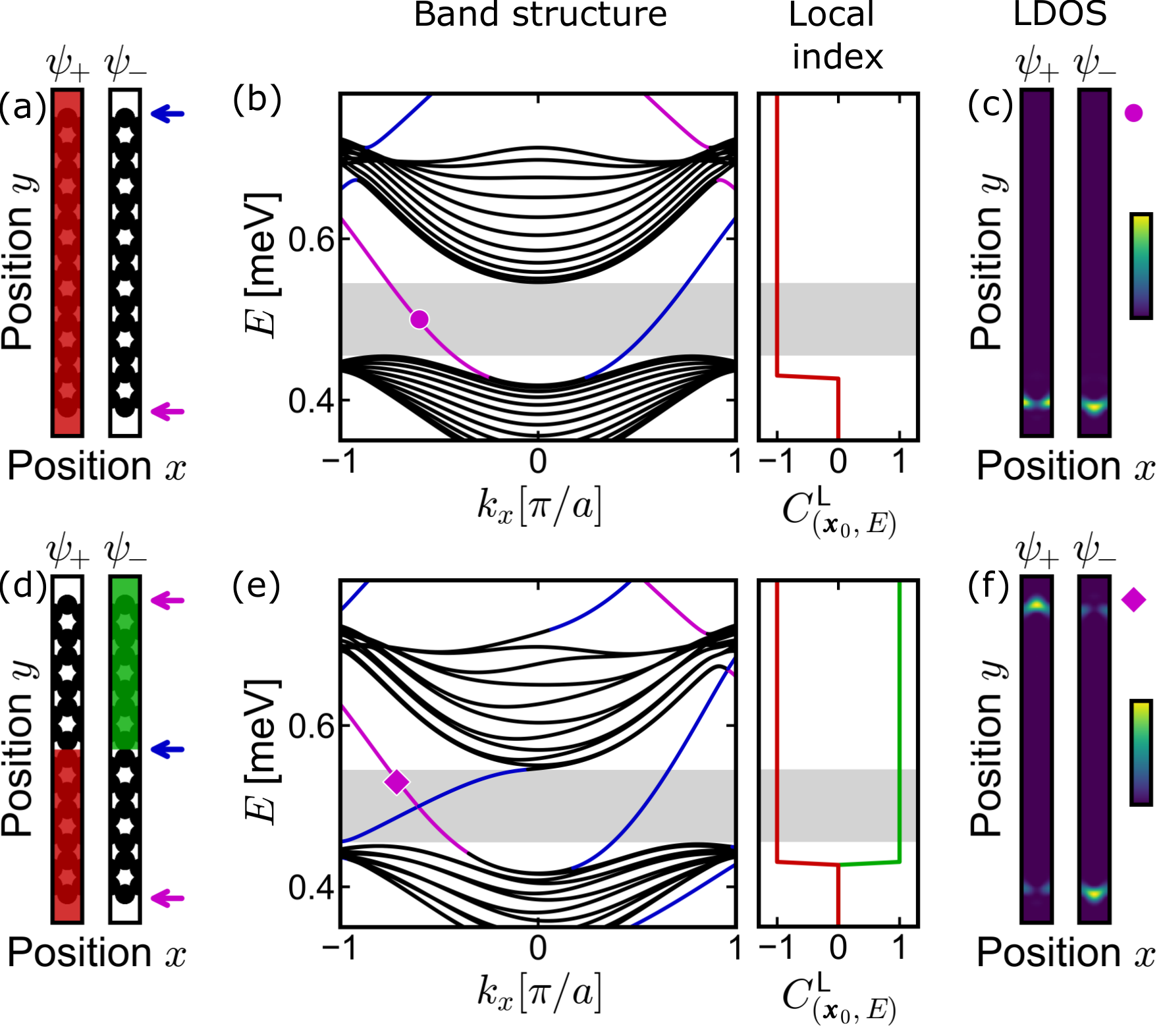}
\caption{
\textbf{Change of the topology due to the spin-dependent blueshift.}
(a) Potential landscape $V(\vec{x})$ of the ribbon polariton lattice for the two spin sectors $\psi_+$, $\psi_-$, arranged in a honeycomb lattice. 
The black and white regions correspond to a potential of $V=6~\unit{\meV}$ and $V=0~\unit{\meV}$, respectively.
The red overlay denotes the blueshift applied on a positive spin sector ($\psi_+$).
(b) Ribbon band structure of (a), and the corresponding local Chern number $C_{(\vec{x}_0,E)}^{\textrm{L}}$, at a the given energy, calculated using the spectral localizer.
In the band structure, the black lines correspond to the bulk modes, and the magenta (blue) lines denotes the topological edge mode dispersion localized at the bottom (top) side of the lattice, as shown by the color-coded arrows in (a).
The gray shaded area indicate the first band gap, which is the energy range of interest. 
(c) Local density of states (LDOS) of the magenta line at $E=0.5~\unit{\meV}$, $k_x = -0.6 [\pi/a]$. 
(d)-(f) Same as (a)-(c) but with the blueshift applied on both spin sectors and on opposite half of the ribbon lattice.
The blue (and magenta) lines in the band structure correspond to the topological edge modes localized at the interface between the blueshifted areas (and at the edges of the lattice).
In (e), the solid red and green lines are the local Chern number calculated inside the $\psi_+$-blueshifted and $\psi_-$-blueshifted regions, respectively.
(f) LDOS of the magenta line at $E=0.53~\unit{\meV}$, $k_x = -0.72 [\pi/a]$.
Parameter values:
lattice constant is $a = 2.95~\unit{\um}$ (center-to-center is $1.7~\unit{\um}$), 
radius of the rods is $1~\unit{\um}$;
$m = 1.3 \times 10^{-4} m_0$ with $m_0$ the free electron mass, 
$\beta_\text{eff} = 0.2~\unit{\meV.\um^2}$,
$\Delta_\text{eff} = 0~\unit{\meV}$;
spin-dependent blueshift energy used is $E_\text{blueshift}=0.3~\unit{\meV}$;
$\kappa = 0.01~\unit{\meV\per\um}$.
}
\label{fig_supp:band_spin}
\end{figure}

Non-trivial topological polariton lattices can be realized solely using a circularly polarized pump~\cite{Bleu2016, Bleu2017, Solnyshkov2018, Sigurdsson2019}.
Using a circularly polarized non-resonant pump will lead to nonlinear interactions only to one of the spin sectors, inducing a blueshift in the corresponding polariton spin sector~\cite{Ferrier2011, Banerjee2021}.
As a result, by redefining the reference polariton energy, an effective Zeeman splitting term $\tilde{\Delta}_\text{eff} = E_\text{blueshift}$ is achieved, known as optical Zeeman splitting, and can be used for inducing non-trivial topology
\begin{equation}
\label{eq:fem_localizer_2d_nh}
B = 
\left(
\begin{array}{cc}
E_\text{blueshift} & 0 \\
0 & 0 \\
\end{array}
\right)
=
\left(
\begin{array}{cc}
\tilde{E}_0 + \frac{1}{2} \tilde{\Delta}_\text{eff} & 0 \\
 & \tilde{E}_0 - \frac{1}{2} \tilde{\Delta}_\text{eff} \\
\end{array}
\right)
,
\end{equation}
with $\tilde{E}_0 = E_\text{blueshift}/2$.
Figure~\ref{fig_supp:band_spin}(a) illustrates this optical Zeeman splitting process.
The landscape of the polariton potential is shown for both spin sectors, where a blueshift is only applied on the $\psi_+$-sector (shown in red shaded area).
Similar to the previous band calculations, the blueshift is manually added to the $\psi_+$-subspace to emulate the positive circular pump, with $\Gamma = 0$. 
The corresponding ribbon band structure (left panel) and the local Chern number at the associated energy in (right panel) is plotted in Fig.~\ref{fig_supp:band_spin}(b), demonstrating the opening of topological band gap with a local Chern number $C_{(\vec{x}_0,E)}^{\textrm{L}} = -1$.
The magenta (blue) lines in the band structure depict the topological edge modes localized at the top (bottom) lattice edge, as shown by the LDOS of the magenta line at $E = 0.5~\unit{\meV}$ in Fig.~\ref{fig_supp:band_spin}(c).
Note that similar results can be realized by applying a blueshift on the $\psi_-$-subspace, with the difference being that the local Chern number will be opposite, $C_{(\vec{x}_0,E)}^{\textrm{L}} = 1$.

With a $\psi_+$-blueshift on one part of the lattice and a $\psi_-$-blueshift on the other part of the lattice, an internal topological interface can be created.
In particular, these blueshifts can be realized by illuminating the sample with a left (or right) circular polarized non-resonant pump in one part of the lattice (or the other), as shown in Fig.~\ref{fig_supp:band_spin}(d) where the red (green) shaded area denotes the induced blueshift on the $\psi_+$ ($\psi_-$) subspace.
Such opposite circular polarized pump configuration will lead to topological modes at the interface between the $C_{(\vec{x}_0,E)}^{\textrm{L}} = -1$ and $C_{(\vec{x}_0,E)}^{\textrm{L}} = 1$ regions of the lattice.
Consequently, a new topological interface is formed with a local Chern number difference of $|\Delta C_{(\vec{x}_0,E)}^{\textrm{L}}| = 2$, resulting in two topological edge modes at the interface [see Fig.~\ref{fig_supp:band_spin}(e)].
Note that the magenta line in the ribbon band structure [see left panel of Fig.~\ref{fig_supp:band_spin}(e)] is doubly degenerate, and its associated LDOS is plotted in Fig.~\ref{fig_supp:band_spin}(f). 
Specifically, the $\psi_+$ ($\psi_-$) spin sector has higher LDOS amplitude on the top (bottom) edge of the lattice. 
%


\subsection{Probing the topology in polariton lattices with spin-dependent blueshift}
\label{sect_supp:localizer_spin}

\begin{figure}[!]
\center
\includegraphics[width=\columnwidth]{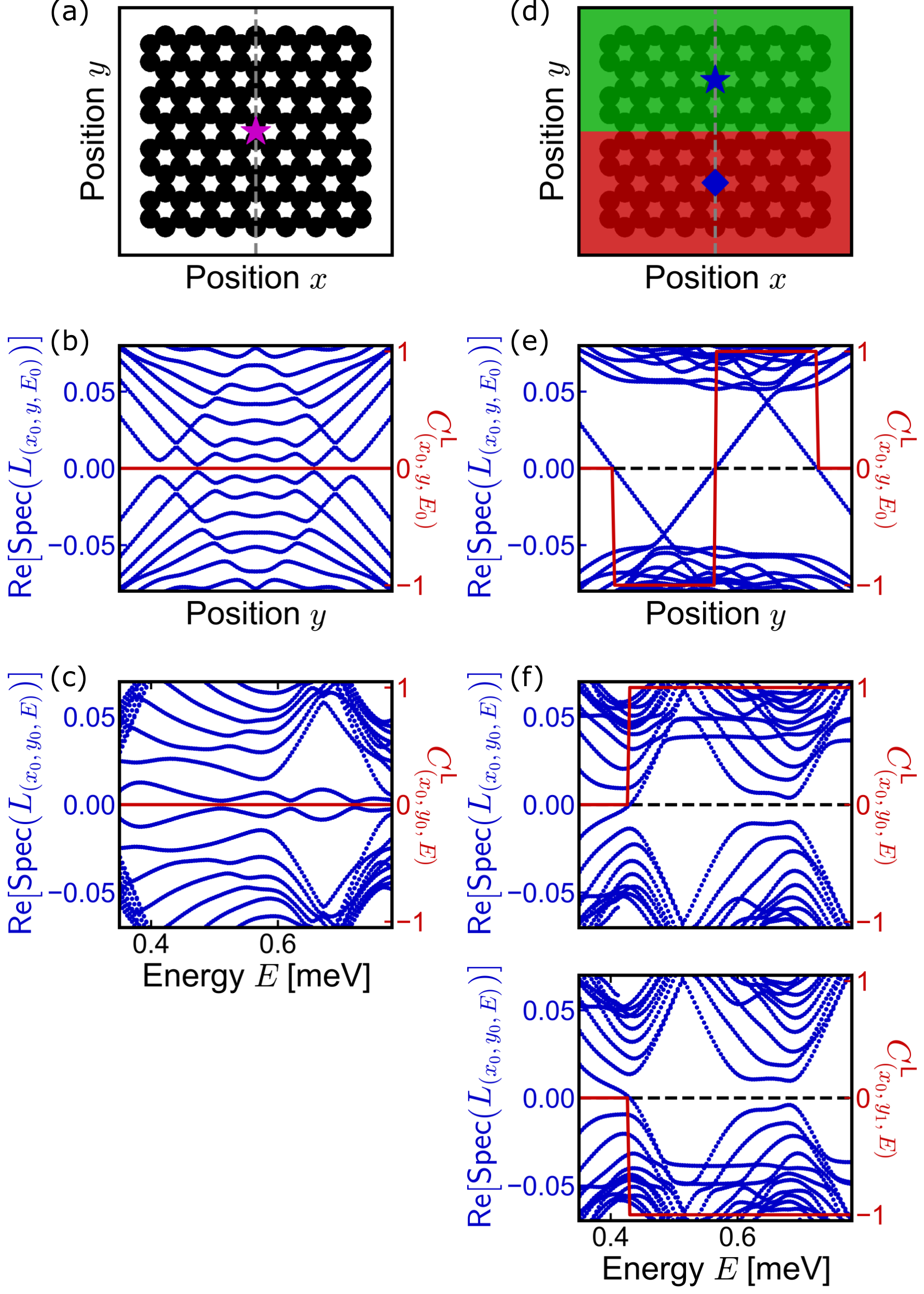}
\caption{
\textbf{Probe of the topology with the spectral localizer for systems with non-resonant circular polarized pump.}
(a) Potential landscape $V(\vec{x})$ of the polariton lattice arranged in a honeycomb lattice. 
The black and white regions correspond to a potential of $V=6~\unit{\meV}$ and $V=0~\unit{\meV}$, respectively.
(b) Eigenvalues of the spectral localizer $\text{Re} \left[ \text{Spec} \left( L_{(x_0,y,E_0)} \right) \right]$ and the local Chern number $C_{(x_0,y,E_0)}^{\textrm{L}}$ along the dashed line in (a) and at $E=0.5~\unit{\meV}$.
(c) Eigenvalues of the spectral localizer $\text{Re} \left[ \text{Spec} \left( L_{(x_0,y_0,E)} \right) \right]$ and the local Chern number $C_{(x_0,y_0,E)}^{\textrm{L}}$ along the energy axis and at the position of the magenta star in (a).
(d)-(f) Same as (a)-(c) but with a red and green overlay on the lattice in (d), depicting the blueshift on the $\psi_+$ and $\psi_-$ spin sectors.
The top and bottom panels of (f) are calculated at the loaction of the blue star ($x_0,y_0$) and blue diamond ($x_0,y_1$), respectively.
Parameter values:
lattice constant is $a = 2.95~\unit{\um}$ (center-to-center is $1.7~\unit{\um}$), 
radius of the rods is $1~\unit{\um}$;
$m = 1.3 \times 10^{-4} m_0$ with $m_0$ the free electron mass, 
$\beta_\text{eff} = 0.2~\unit{\meV.\um^2}$,
$\Delta_\text{eff} = 0~\unit{\meV}$;
spin-dependent blueshift energy used is $E_\text{blueshift}=0.3~\unit{\meV}$;
$\kappa = 0.01~\unit{\meV\per\um}$.
}
\label{fig_supp:localizer_spin}
\end{figure}

Here, we present how topology has been studied for a polaritonic system with no external magnetic, and with a blueshift on a single spin sector [see for example Figs.~\ref{fig_supp:band_spin}(b),(e) in the main text], with $\Gamma = 0$.
The topology in the polariton lattice is characterized directly from the continuous model with the spectral localizer, similar to what is discussed in Sect.~\ref{sect_supp:localizer}.

In the absence of non-resonant pump, namely without any added blueshift, the system is topologically trivial, as shown in Fig.~\ref{fig_supp:localizer_spin}(b),(c).
In particular, even though the system is gapless because of the lack of external magnetic field, and lacks time-reversal symmetry breaking, its topology can still be calculated with the spectral localizer and the local Chern number.
Figure~\ref{fig_supp:localizer_spin}(b) shows that the local Chern number is zero along the dashed line in Fig.~\ref{fig_supp:localizer_spin}(a) at $E_0 = 0.5~\unit{\meV}$, and Figure~\ref{fig_supp:localizer_spin}(c) indicates that the system is trivial at least up to $1~\unit{\meV}$.

With opposite non-resonant circular polarized pumps, as depicted with green and red overlays in Fig.~\ref{fig_supp:localizer_spin}(d), opposite optical Zeeman effects arise in different regions of the lattice.
Figure~\ref{fig_supp:localizer_spin}(e) plots the spectral flow of $L_{(x_0,y,E_0)}$ and the local Chern number $C_{(x_0,y,E_0)}^{\textrm{L}}$ along the dashed in Fig.~\ref{fig_supp:localizer_spin}(d) at $E_0 = 0.5~\unit{\meV}$, revealing the system is topologically non-trivial and that the upper and lower halves of the lattice possess opposite local Chern numbers.
Looking at the topology at the location of the blue star (blue diamond) [see Fig.~\ref{fig_supp:localizer_spin}(d)], the top (bottom) panel of Fig.~\ref{fig_supp:localizer_spin}(f) indicates the system is non-trivial for some energy range starting from around $0.4~\unit{\meV}$ with local Chern number $C_{(x_0,y_0,E)}^{\textrm{L}} = 1$ ($C_{(x_0,y_1,E)}^{\textrm{L}} = -1$). 

The local Chern numbers shown in Fig.~\ref{fig_supp:band_spin} correspond to the calculated local Chern number along the energy axis in Figs.~\ref{fig_supp:localizer}(c),(f).


\subsection{Topological routing}
\label{sect_supp:reconfig_spin}

\begin{figure}[t]
\center
\includegraphics[width=\columnwidth]{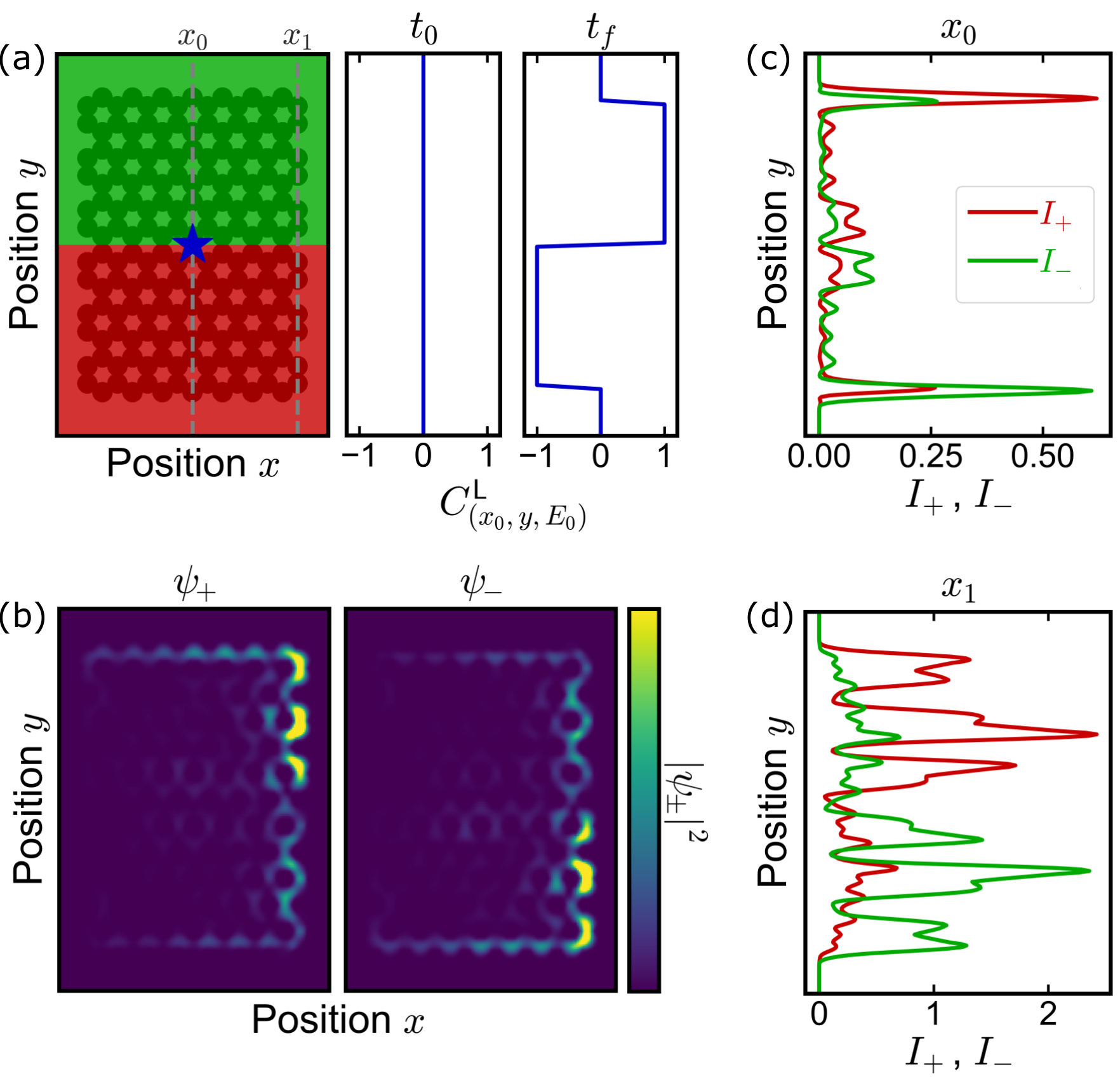}
\caption{
\textbf{Reconfigurable topological routing with non-resonant circular polarized pumping.}
(a) Potential landscape $V(\vec{x})$ of the polariton lattice arranged in a honeycomb lattice, and the local Chern number $C_{(x_0,y,E_0)}^{\textrm{L}}$ along the $x_0$ gray dashed line and at $E_0=0.5~\unit{\meV}$, calculated using the spectral localizer before the pump starts ($t_i$) and at the final time ($t_f$).
The red and green overlays depict the positive ($S_{\text{pump},+}$) and negative ($S_{\text{pump},-}$) circular polarized pump, respectively.
The blue star indicate the position of the probe source.
(b) Snapshot of the intensity of the polariton for each spin sectors $|\psi_+|^2$, $|\psi_-|^2$ at $t = 455 [\hbar] \approx 300~\unit{\ps}$.
(c)-(d) Integrated intensity of each spin sector over a width of $1~\unit{\um}$ in the y-direction, at the positions given by the red and green crosses in (a), along the $x_0$ and $x_1$ gray dashed line in (a).
The parameter values for the Hamiltonian are the same as in Fig.~\ref{fig_supp:band_spin}.
Dynamical parameter values:
$\gamma_c = 0.05~\unit{\per\ps}$, 
$\gamma_r = 1.5 \gamma_c$, 
$g_c = 5 \times 10^{-3}~\unit{\meV.\um^2}$,
$g_r = 10 \times 10^{-3}~\unit{\meV.\um^2}$,
$R = 3 \times 10^{-4}~\unit{\ps^{-1}\um^2}$;
$S_{0,\text{pump}} = 2.3~\unit{\ps^{-1}\um^{-2}}$,
$S_{0,\text{probe}} = 0.5~\unit{\ps^{-1}\um^{-2}}$;
$\hbar \omega_\text{s} = 0.5~\unit{\meV}$,
$k_\text{s} = 0.5 [\pi/a]$;
$\kappa = 0.01~\unit{\meV\per\um}$. 
}
\label{fig_supp:reconfig_spin}
\end{figure}

To illustrate topological routing using the optical Zeeman effect, we consider the dynamic response of the system shown in Fig.~\ref{fig_supp:reconfig_spin}(a). 
The red and green overlays on the polariton potential denote the pump pattern with positive and negative circular polarized non-resonant pumps, inducing a $\psi_+$-blueshift and $\psi_-$-blueshift respectively.
In particular, the dynamics of system is calculated using the modified rate equations [Eqs.~\eqref{eq:polariton_rate}-\eqref{eq:reservoir_rate}] from the continuous model. 
Additionally, two slowly increasing and circularly polarized non-resonant pumping sources are used, as depicted in Fig.~\ref{fig_supp:reconfig_spin} with the red and green overlays, each with maximum amplitudes $S_{0,\text{pump}} = 2.3~\unit{\ps^{-1}\um^{-2}}$ below the condensate threshold; 
and a resonant Gaussian source with amplitude $S_{0,\text{probe}} = 0.5~\unit{\ps^{-1}\um^{-2}}$, located at the blue star [see Fig.~\ref{fig_supp:reconfig_spin}(a)], is used to excite the topological modes.
Note that, without loss of generality, the spin relaxation is here not included in our model~\cite{Banerjee2021, Ohadi2012}, although the main effect will be to increase the non-resonant pump powers to achieve the same desired optical Zeeman strength. 
See the Section~\ref{sect_supp:spin_relax} for additional information.

The resulting dynamics of the change in real-space topology from the difference in the local Chern number [see the right panels of Fig.~\ref{fig_supp:reconfig_spin}(a)] demonstrates the emergence of a topological interface in the interior of the lattice and thus the existence of topological modes.
At the initial time ($t_0$), before the lattice is illuminated, the topology of the system is trivial with local Chern number being $C_{(x_0,y,E_0)}^{\textrm{L}} = 0$ all along the gray dashed and at $E_0=0.5~\unit{\meV}$.
At the final simulated time ($t_f$), under circular polarized illumination, the system is topologically non-trivial with local Chern numbers being $C_{(x_0,y,E_0)}^{\textrm{L}} = -1$ and $C_{(x_0,y,E_0)}^{\textrm{L}} = 1$ in the lower and upper half of the lattice, respectively, in according with the polarized pump patterns.
Figure~\ref{fig_supp:reconfig_spin}(b) shows a time snapshot of the intensities $|\psi_+|^2$, $|\psi_-|^2$ for each spin sector, demonstrating the excitation and propagation of the topological mode along the newly formed topological interface.
After reaching the right edge of the lattice, the topological mode is split into an upward and downward propagating topological mode along the outer topological interface from the lattice edge.

Remarkably, at the bifurcation of different topological interfaces, a finite spin separation occurs.
In particular, when the excited topological modes reached the right edge of the lattice and split into upward and downward propagating topological modes [see Fig.~\ref{fig_supp:reconfig_spin}(b)], $\psi_+$ polariton states are dominant in the upper half of the lattice, and vice versa for the $\psi_-$ states. 
This spin separation is quantified by looking at the integrated intensities $I_+$, $I_-$ over finite width of $3~\unit{\um}$ centered around $x_0$ and $x_1$ [see dashed line sin Fig.~\ref{fig_supp:reconfig_spin}(a)], as plotted in Figs.~\ref{fig_supp:reconfig_spin}(c)-(d). 
At $x_1$ [see Fig.~\ref{fig_supp:reconfig_spin}(d)], on the right lattice edge, the spin up intensity $I_+$ is predominant in the upper half lattice edge, while the spin down $I_-$ is higher on the lower half.
At $x_0$ [see Fig.~\ref{fig_supp:reconfig_spin}(c)], the topological edge mode along the top (or bottom) has higher $I_+$ (or $I_-$).
Note that similar topological mode propagation can be achieved when combined with a linearized polarized pump, inducing an internal trivial gapped region and topological interface, akin to the outer edge of the lattice.
Consequently, using a linearly polarized pump in addition to particular circularly polarized pump patterns provides a possible control mechanism for topological modes, as well as over the position where the modes bifurcate and the spin separation arises.

In Figure~\ref{fig_supp:reconfig_spin}, the pump is given by
\begin{equation}
S_{\text{pump}}
= 
\begin{cases}
S_{0,\text{pump}} \frac{1}{1 + e^{-\frac{t-t_0}{2\tau}}} 
\left(
\begin{array}{c}
\mathbf{1} \\[1.ex] 
\mathbf{0}
\end{array}
\right)
& \text{for $y$ in lower half}
\\[3.ex] 
S_{0,\text{pump}} \frac{1}{1 + e^{-\frac{t-t_0}{2\tau}}} 
\left(
\begin{array}{c}
\mathbf{0} \\[1.ex] 
\mathbf{1}
\end{array}
\right)
& \text{for $y$ in upper half}
\end{cases}
,
\end{equation}
with 
$t_0 = 10 [\hbar]$, 
$\tau_t = 0.9 [\hbar]$,
and the resonant probe is a Gaussian source, located at $(x_\text{s}, y_\text{s}) = (0, 0)$ [see blue star in Fig.~\ref{fig_supp:reconfig_spin}(a)] and centered at a time $t_0 = 300 [\hbar]$,
\begin{equation}
\begin{split}
S_{\text{probe}}
& = S_{0,\text{probe}} e^{ -\frac{(x_\text{s}-x)^2+(y_\text{s}-y)^2}{2\tau_{xy}^2} } e^{ -\frac{(t-t_0)^2}{2\tau_t^2} } \\
& \qquad \qquad \times 
e^{-i \omega_\text{s} t} e^{i k_\text{s} x} 
\left(
\begin{array}{c}
\mathbf{0} \\[1.ex] 
\mathbf{1}
\end{array}
\right)
,
\end{split}
\end{equation}
with 
$\tau_{xy} = 1~\unit{\um}$, 
$\tau_t = 40 [\hbar]$.
%


\section{Effect of spin relaxation for the optical Zeeman splitting}
\label{sect_supp:spin_relax}

\begin{figure}[!]
\center
\includegraphics[width=\columnwidth]{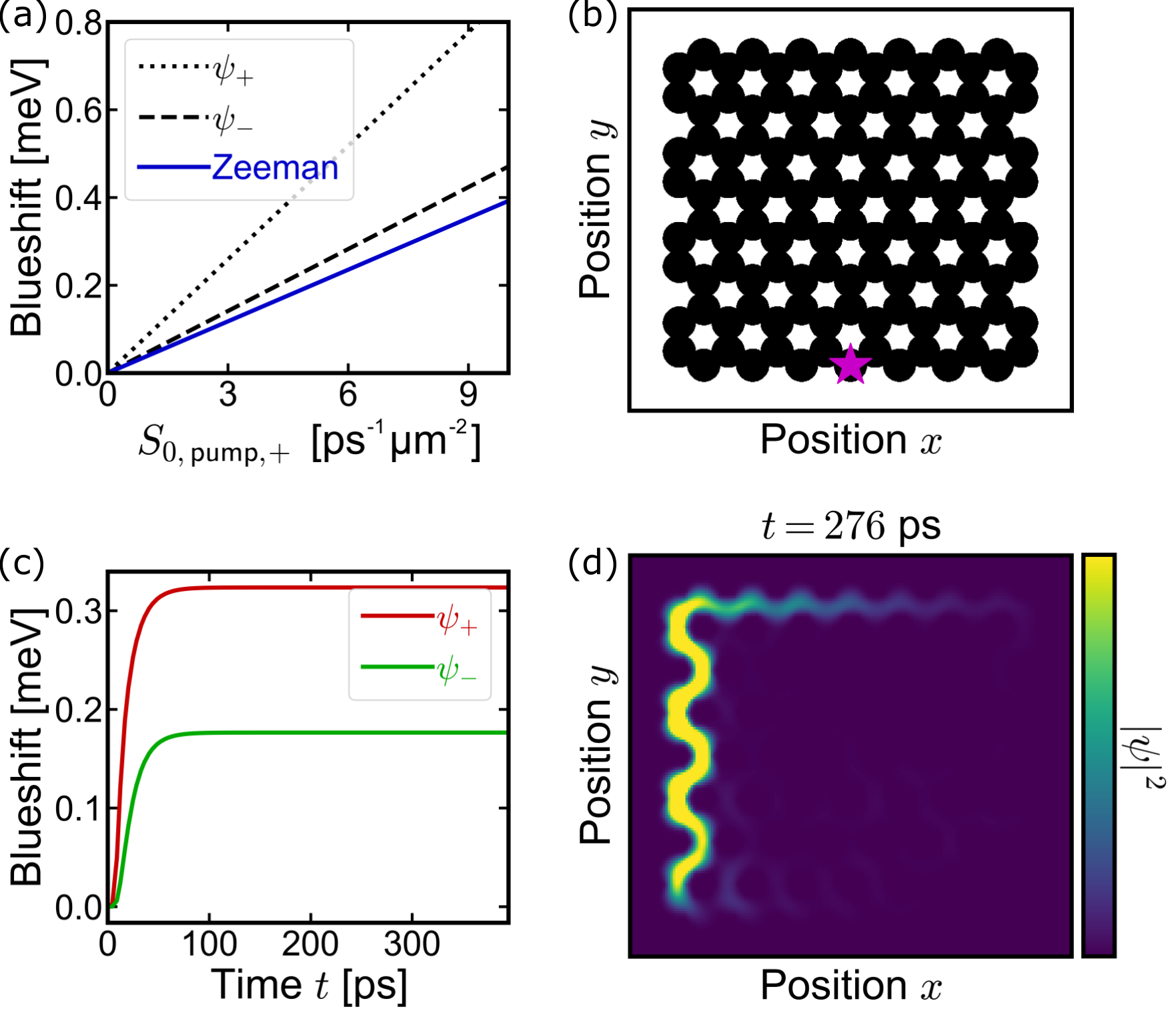}
\caption{
\textbf{Optical Zeeman effect with spin relaxations.}
(a) Theoretical blueshift from the steady state solution of the $\psi_+$ and $\psi_-$ density exciton reservoir in the rate equations [Eq.~\eqref{eq_supp:reservoir_rate}], with $S_{0,\text{pump},-} = 0~\unit{\ps^{-1}\um^{-2}}$. 
Blue line is the effect of the Zeeman splitting, calculated from the difference between the $\psi_+$ and $\psi_-$ blueshift.
(b) Potential landscape $V(\vec{x})$ of the polariton lattice arranged in a honeycomb lattice.
The magenta star indicate the position of the probe source.
(c) Temporal evolution of the spin-dependent blueshift. 
(d) Snapshot of the total intensity of the polariton $|\psi|^2 = |\psi_+|^2 + |\psi_-|^2$.
The parameter values for the Hamiltonian are the same as in Fig.~\ref{fig_supp:localizer_spin}.
Dynamical parameter values:
$\gamma_c = 0.05~\unit{\per\ps}$, 
$\gamma_r = 1.5 \gamma_c$, 
$g_c = 5 \times 10^{-3}~\unit{\meV.\um^2}$,
$g_r = 10 \times 10^{-3}~\unit{\meV.\um^2}$,
$R = 3 \times 10^{-4}~\unit{\ps^{-1}\um^2}$,
$J = 0.09~\unit{\per\ps}$;
$S_{0,\text{pump},+} = 7.5~\unit{\ps^{-1}\um^{-2}}$,
$S_{0,\text{pump},-} = 0~\unit{\ps^{-1}\um^{-2}}$,
$S_{0,\text{probe},\pm} = 0.5~\unit{\ps^{-1}\um^{-2}}$;
$\hbar \omega_\text{s} = 0.86~\unit{\meV}$,
$k_\text{s} = 0.5 [\pi/a]$.
}
\label{fig_supp:spin_relax}
\end{figure}

As the optical Zeeman effect relies on the spins of the exciton reservoir $n_r(\vec{x},t) = [n_{r,+}(\vec{x},t), n_{r,-}(\vec{x},t)]$, this section studies the effect of considering spin relaxations in the rate equations.
The spin relaxation is included by considering the couplings $J$ between the different spins in the exciton reservoir.
The rate equations are therefore slightly modified as~\cite{Banerjee2021}
\begin{align}
\label{eq_supp:polariton_rate}
\begin{split}
i \hbar \frac{\partial}{\partial t} \vec{\psi} 
& = H_0 \vec{\psi} - i \hbar \frac{\gamma_c}{2} \vec{\psi} + g_c \left| \vec{\psi} \right| ^2 \vec{\psi} \\
& \qquad \qquad + \left( g_r + i \hbar \frac{R}{2} \right) n_r \vec{\psi} + S_{\text{probe}},
\end{split}
\\
\label{eq_supp:reservoir_rate}
\begin{split}
\frac{\partial}{\partial t} n_{r,\pm} 
& = - \left( \gamma_r + R \left| \vec{\psi_\pm}  \right| ^2 \right) n_{r,\pm} + J \left(  n_{r,\mp} - n_{r,\pm} \right) \\
& \qquad \qquad + S_{\text{pump},\pm},
\end{split}
\end{align}
where 
$H_0$ is the Hamiltonian and denotes the kinetic energy and the couplings between of the polaritons, 
$\gamma_c$ and $\gamma_r$ are the relaxation rates for the polariton state and exciton reservoir, 
$g_c$ and $g_r$ are the polariton-polariton and polariton-exciton interaction strengths, 
$R$ is the amplification rate of the polariton state due to stimulated scattering of polariton from the reservoir,
$S_\text{probe}(\vec{x}, t)$ is the resonant probe for exciting the polaritons,
and $S_\text{pump}(\vec{x}, t)$ is the non-resonant pump for exciting the free carriers.

One of the main consequences of spin relaxation is the pump power used to achieved a given blueshift on the spin sectors.
Indeed, from Eq.~\eqref{eq_supp:reservoir_rate}, the steady states of the density of the exciton reservoir reads
\begin{align}
\label{eq_supp:n2}
\begin{split}
n_{r,-,\text{th}}
& = \frac{J}{\gamma_r(\gamma_r+2J)} S_{\text{pump},+} + \frac{\gamma_r+J}{\gamma_r(\gamma_r+2J)} S_{\text{pump},-}
,
\end{split}
\\
\label{eq_supp:n1}
\begin{split}
n_{r,+,\text{th}} 
& = \frac{1}{\gamma_r+J} S_{\text{pump},+} + \frac{J}{\gamma_r+J} n_{r,-,\text{th}}
.
\end{split}
\end{align}
As a result, by redefining the reference polariton energy, an effective Zeeman splitting term $\tilde{\Delta}_\text{eff} = E_{\text{blueshift},+} - E_{\text{blueshift},-}$ is achieved and the the overall system is blueshifted by $\tilde{E}_0 = (E_{\text{blueshift},+} + E_{\text{blueshift},-})/2$
\begin{equation}
\label{eq:fem_localizer_2d_nh}
\begin{split}
B 
& = 
\left(
\begin{array}{cc}
E_{\text{blueshift},+} & 0 \\
0 & E_{\text{blueshift},-} \\
\end{array}
\right)
\\[1.ex]
& = 
\left(
\begin{array}{cc}
\tilde{E}_0 + \frac{1}{2} \tilde{\Delta}_\text{eff} & 0 \\
 & \tilde{E}_0 - \frac{1}{2} \tilde{\Delta}_\text{eff} \\
\end{array}
\right)
.
\end{split}
\end{equation}
Thus, injecting single spin excitons will lead to a finite density of the other exciton spin as well, meaning that higher pump powers are required to achieve the same amount of blueshift as in the case of neglected spin relaxation.
According to Fig.~\ref{fig_supp:spin_relax}(a), a blueshift of $\tilde{\Delta}_\text{eff} = 0.3~\unit{\meV}$ needs a pump power of $S_{0,\text{pump},+} = 7.5~\unit{\ps^{-1}\um^{-2}}$, $S_{0,\text{pump},-} = 0~\unit{\ps^{-1}\um^{-2}}$, compared to $S_{0,\text{pump}} = 2.3~\unit{\ps^{-1}\um^{-2}}$ in Sect.~\ref{sect_supp:reconfig_spin} when neglecting the spin relaxation.

Figures~\ref{fig_supp:spin_relax}(b)-(d) shows the dynamics of the system using the modified rate equations [Eqs.\eqref{eq_supp:polariton_rate}-\eqref{eq_supp:reservoir_rate}].
In particular, a non-resonant positive circularly polarized pump is illuminating the whole lattice (the overlay of the pump is not shown here), leading to the injection of $\psi_+$-spin excitons in the reservoir.
The temporal evolution of the blueshift derived from the $\psi_+$ and $\psi_-$ excitons is shown in Fig.~\ref{fig_supp:spin_relax}(c), demonstrating the $\psi_+$ excitons indeed relaxed into $\psi_-$.
A resonant Gaussian source is used to excite the topological edge mode once the steady state of the  exciton reservoir is reached, and a snapshot of the excited topological mode is plotted in Fig.~\ref{fig_supp:spin_relax}(d), showing that the edge mode is propagating along the edge without being back reflected, as expected.


\end{document}